\documentclass[journal]{IEEEtran}
\usepackage{cite}
\usepackage{amsmath,amssymb,amsfonts}
\usepackage{graphicx}
\usepackage{amssymb}
\usepackage{pifont}
\usepackage{lipsum}
\usepackage{textcomp}
\usepackage{multirow}
\usepackage{dblfloatfix}
\usepackage{algorithmic,eqparbox,array}
\usepackage{color}
\usepackage[super]{nth}
\usepackage{algorithm}
\usepackage{comment}
\usepackage{multicol}
\usepackage{multirow}
\usepackage{booktabs}
\usepackage{subcaption}
\usepackage{graphicx}
\usepackage{hyperref}
\usepackage{forest}
\usepackage{tikz-qtree}
\usepackage{algorithmic}
\usepackage{cleveref}
\usepackage{balance}
\usepackage{tikz}
\usetikzlibrary{shapes,arrows}
\usepackage[latin1]{inputenc}
\usepackage{commath}
\usepackage{xurl}
\newcommand{\centered}[1]{\begin{tabular}{l} #1 \end{tabular}}

\newcommand{\etal}{{\em et al.}}
\newcommand{\ie}{{\em i.e.}}
\newcommand{\eg}{{\em e.g.}}
\newcommand{\changes}[1]{\textcolor{black}{#1}}
\newcommand{\changess}[1]{\textcolor{black}{#1}}
\ifCLASSINFOpdf
\else
\fi

\begin{document}
\title{Trojan Playground: A Reinforcement Learning Framework for Hardware Trojan Insertion and Detection}
\author{Amin~Sarihi,~\IEEEmembership{Student Member,~IEEE,}
Ahmad Patooghy,~\IEEEmembership{Member,~IEEE,}
Peter Jamieson,~\IEEEmembership{Member,~IEEE,}
and Abdel-Hameed~A.~Badawy,~\IEEEmembership{Senior~Member,~IEEE}
\thanks{Amin Sarihi is a Ph.D.\ candidate at the Klipsch School of Electrical and Computer Engineering, New Mexico State University, Las Cruces, NM, 88001 USA. E-mail: sarihi@nmsu.edu}
\thanks{Prof. Ahmad Patooghy is an assistant professor at the Department of Computer Systems Technology at North Carolina A\&T State University, Greensboro, NC, 27411, USA. E-mail: apatooghy@ncat.edu}
\thanks{Prof. Peter Jamieson is an associate professor at the Electrical and Computer Engineering Department at Miami University, Oxford, OH, 45056, USA. E-mail: jamiespa@miamioh.edu}
\thanks{Prof. Abdel-Hameed A. Badawy is an Associate Professor at the Klipsch School of Electrical and Computer Engineering, New Mexico State University, Las Cruces, NM, 88001 USA. E-mail: badawy@nmsu.edu}
\vspace{-10mm}
}




\maketitle


\begin{abstract}
Current Hardware Trojan (HT) detection techniques are mostly developed based on a limited set of HT benchmarks. Existing HT benchmark circuits are generated with multiple shortcomings, \ie, i) they are heavily biased by the designers' mindset when created, and ii) they are created through a one-dimensional lens, mainly the signal activity of nets. We introduce the first automated Reinforcement Learning (RL) HT insertion and detection framework to address these shortcomings. In the HT insertion phase, an RL agent explores the circuits and finds locations best for keeping inserted HTs hidden. On the defense side, we introduce a multi-criteria RL-based HT detector that generates test vectors to discover the existence of HTs. Using the proposed framework, one can explore the HT insertion and detection design spaces to break the limitations of human mindset and benchmark issues, ultimately leading toward the next generation of innovative detectors. We demonstrate the efficacy of our framework on ISCAS-85 benchmarks, provide the attack and detection success rates, and define a methodology for comparing our techniques.
\end{abstract}

\begin{IEEEkeywords}
Hardware Trojan, Hardware Security, Reinforcement Learning, Open-Source.
\end{IEEEkeywords}

%
\IEEEpeerreviewmaketitle

\section{Introduction}\label{sec:intro}
\begin{figure}[!h]
  \centering
  \includegraphics[width=.49\textwidth]{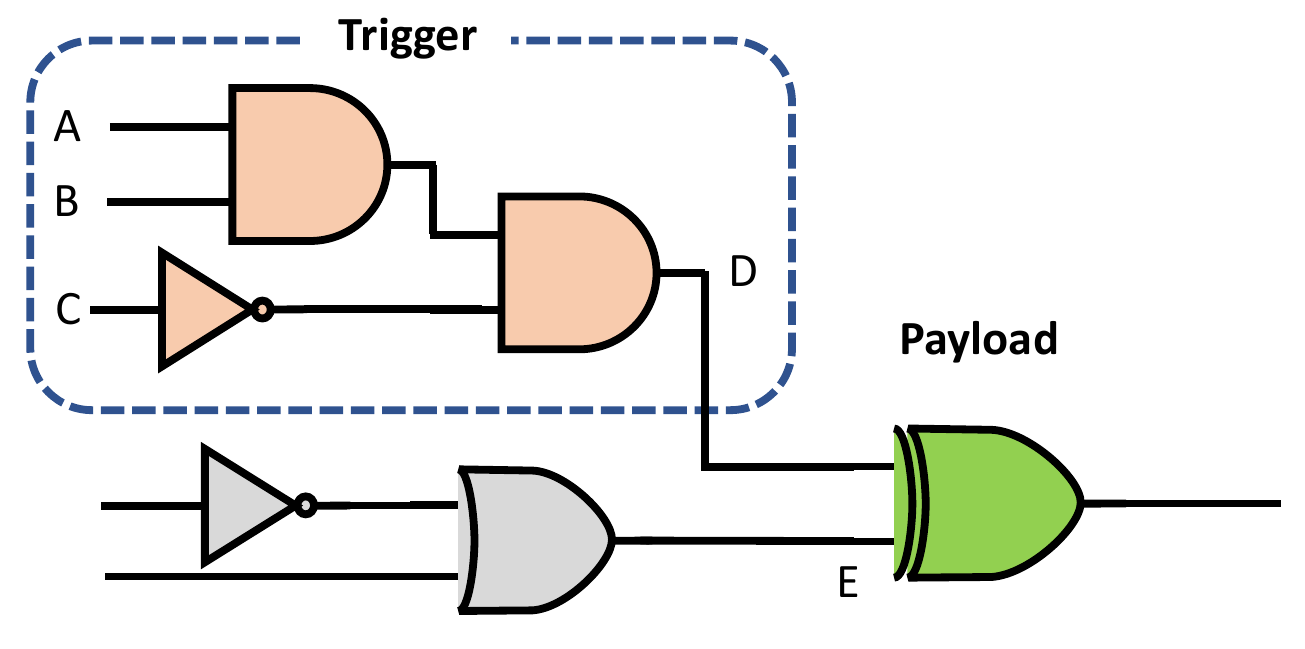}
  \caption{An HT with a trigger and payload. Whenever $A=1$, $B=1$, and $C=0$, the trigger is activated ($D=1$) and the XOR payload inverts the value of $E$. }
  \label{trigger_payload}
\end{figure}

Per a DoD report~\cite{securing} released in $2022$, $88\%$ of the production and $98\%$ of the assembly, packaging, and testing of microelectronic chips are performed outside of the US. The growing multi-party production model has significantly raised security concerns about malicious modifications in the design and fabrication of chips, \ie, Hardware Trojan (HT) insertion. \textit{HTs} are defined as any design or manufacturing violations in an integrated circuit (IC) concerning the intent of the IC. Upon activation, an HT may lead to erroneous outputs (\eg, Figure~\ref{trigger_payload}) and possibly leak of information~\cite{pan2021automated}. According to the adversarial model introduced by Shakya~\etal~\cite{shakya2017benchmarking}, HTs can be inserted into target ICs according to the following scenarios:

\begin{itemize}
    \item  Design source code or netlist can be infected with HTs by compromised employees.
    \item Third-party intellectual properties (IPs) like processing cores, memory modules, I/O components, and network-on-chip~\cite{sarihi2021survey} are often purchased and incorporated into a design to speed up time-to-market and lower design expenses. However, integrating IPs from untrusted vendors can pose a risk to the security and integrity of the IC.
    \item An untrusted foundry may reverse-engineer the GDSII physical layout to obtain the netlist and insert HTs inside them. 
    \item Malicious third-party CAD tools may also insert HTs into designs
\end{itemize}

Researchers have been mostly using established benchmarks reported by Shakya~\etal and Salmani~\etal~\cite{shakya2017benchmarking,salmani2013design} as a reference to study the impact of HTs\footnote{The benchmarks are available on Trust-Hub~\cite{trusthub}.}. Subsequently, various HT detection approaches have been developed based on these benchmarks over the past decade~\cite{salmani2016cotd,hasegawa2017trojan, sebt2018circuit,gohil2022deterrent}. Despite the valuable effort to create HT benchmarks for the community, these benchmarks are limited in size and variety needed to push detection tools into more realistic modern scenarios. For instance, the small set of benchmarks makes it hard to leverage and train machine learning (ML) HT detectors, where more training data negatively impacts classification accuracy. Some research studies have tried to alleviate this problem by using techniques to shuffle data for ML-based detectors, \eg, the leave-one-out cross-validation method~\cite{hasegawa2017trojan}; however, it does not solve the problem entirely. 
The existing HT benchmarks also suffer from an inherent human bias in the insertion phase since they are tightly coupled with the designer's mindset. For instance, the HT benchmarks in~\cite{cruz2018automated} only consider signal activity for HT insertion, \ie, HTs are randomly inserted into a pool of available rare nets of the circuit. The flaws in the insertion phase simplify the problem's complexity, leading security researchers to develop HT detectors finely tuned to flawed scenarios~\cite{lyu2020scalable,gohil2022deterrent}. In contrast, adversaries devise new HT attacks that combine different ideas where detectors fall short of exposing them. Another equally important problem in this domain is having almost no HT detectors publicly available. This deprives other researchers of accessing these tools and imposes a considerable latency for newcomers to hardware security. 

This work attempts to move this research space forward by developing next-generation HT insertion and detection methods based on reinforcement learning.
The developed RL-based HT insertion tool creates new HT benchmarks according to the criteria passed to the tool by the user. The insertion criteria is an RL rewarding function modified by a user that relies on the RL agent to insert HTs into designs automatically. The netlist is considered an environment in which the RL agent tries to insert HTs to maximize a gained reward. The rewarding scheme of the proposed insertion tool is tunable, which can push the agent toward a specific goal in the training session. Our insertion tool is a step towards preparing the community for future HTs inserted by non-human agents, \eg, AI agents. We also propose an RL-based HT detector with a tunable rewarding function that helps detect inserted HTs based on various strategies. To explore this space, we have studied three different detection rewarding functions for the RL detector agent. The agent finds test vectors yielding the highest rewards per each reward function. Then, the generated test vectors activate and find HTs in the IC. The test engineer passes the test vectors to the chip and monitors the output for deviations from the golden model.

Our proposed toolset enables the researchers to experience HT insertion and detection within a unified framework. The framework only requires users to set the parameters to insert and detect HTs without human intervention. There have been previous efforts to automate the HT insertion and detection process~\cite{2018FyrbiakHAL,cruz2018automated,yu2019improved}; however, they need an intermediate effort hindering us from creating a vast quantity of HTs (more explanation in Section~\ref{sec:background}).

Similar to several previous work~\cite{gohil2022deterrent,chakraborty2009mero,pan2021automated,gohil2022attrition,schulman2017proximal}, this paper's threat model assumes that the perpetrator is capable of inserting HTs into a design's netlist. The netlist can be obtained through state-of-the-art reverse-engineering techniques in the foundry, and HT triggers are constructed and placed in the design layout. On the defense side, we assume a security engineer receives a post-silicon hard IP that may or may not contain malicious HTs. Using a golden model, the security engineer generates a set of minimal test vectors to activate as many HTs as possible. The test engineer does not know the insertion criteria; however, they generate test vectors based on multiple insertion mentalities. If the output(s) of the design-under-test deviate(s) from the golden model, it can insinuate malicious behavior.

We make the following contributions in the paper with respect to our previous publications~\cite{sarihi2022hardware, sarihi2023multi}:

\begin{itemize}
    \item We developed a tunable RL-based HT insertion tool free of human bias, capable of automatic HT insertion and creating a large population of valid HTs for each design
    \item We introduce a tunable RL-based multi-criteria HT detection tool that helps a security engineer to better prepare for different HT insertion strategies.  
    \item We introduce and use a generic methodology to compare HT detectors fairly. The methodology is based on the confidence value metric that helps the security engineer select the proper detector based on the chip's application and security requirements. 
\end{itemize}

Our results show that our developed detection tool with all three detection approaches has an average $90.54$\% detection rate for our HT-inserted benchmarks. We compare these detection results to existing state-of-the-art detection methods and show how our techniques find previously unidentifiable HTs. As we believe that HT detection will be implemented as a variety of detection strategies, the uniquely identified HTs suggest that our detection techniques and framework are important contributions to this space. 

The remainder of this paper is organized as follows: Section~\ref{sec:background} reviews the related work and explains the fundamentals of RL. The mechanics of our proposed HT insertion and detection approaches are presented in Sections~\ref{sec:insertion} and~\ref{sec:detection}, respectively. We introduce our HT comparison methodology in Section~\ref{sec:detection_metric}. Section~\ref{sec:Result} demonstrates the experimental results, and Section~\ref{sec:Conc} concludes the paper.

\section{Related work}
\label{sec:background}
This section summarizes the previous studies in HT insertion and detection.

\subsection{Hardware Trojan Insertion and Benchmarks}
\label{subsec:insertion_background}

The first attempts to gather benchmarks with hard-to-activate HTs were made by Shakya~\etal~and  Salmani~\etal~\cite{shakya2017benchmarking,salmani2013design}. A set of $96$ trust benchmarks with different HT sizes and configurations are available at Trust-Hub~\cite{trusthub}. While these benchmarks are a valuable contribution to the research community, they have three drawbacks:

\begin{enumerate}
    \item  The limited number of Trojan circuits represents only a subset of the possible HT insertion landscape in digital circuits, which hampers the ability to develop diverse HT countermeasures, 
    \item They lack incorporating state-of-the-art Trojan attacks, and 
    \item They fail to populate a large enough HT dataset required for ML-based HT detection.
\end{enumerate}

\changes{Krieg \cite{krieg2023reflections} investigates the practicality of the Trusthub benchmark for hardware security study from 5 different perspectives: Correctness, Maliciousness, Stealthiness, Persistence, and Effectiveness. The paper lists nine main flaws that undermine the feasibility of Trusthub for security evaluations, including  pre-/post-synthesis simulation mismatch, unsatisfiable trigger conditions, incorrect original designs, and buggy wiring. The paper shows that out of the 83 benchmarks, only three hold all the properties, and the rest fail in at least one or more studied aspects. }

Various approaches to insert HTs have been attempted. Jyothi~\etal~\cite{jyothi2017taint} proposed a tool called TAINT for automated HT insertion into FPGAs at the RTL level, gate-level netlist, and post-map netlist. The tool also allows the user to insert HTs in FPGA resources such as Look-Up Tables (LUTs), Flip Flops (FFs), Block Random Access Memory (BRAM), and Digital Signal Processors (DSP). 
Despite the claimed automated process, the user is expected to select the trigger nets based on suggestions made by the tool. The results section shows that the number of available nodes in post-map netlists drops significantly, leaving less flexibility for Trojan insertion compared to RTL codes.

Reverse engineering tools can also identify security-critical circuitry in designs that can direct attackers to insert efficient HTs. Wallat~\etal~\cite{wallat2017look}~ introduced HAL, a gate-level netlist reverse engineering tool that offers offensive reverse engineering strategies and defensive measures, such as developing arbitrary Trojan detection techniques. The authors believe that adversaries are more likely to insert HTs through reverse engineering techniques and are less likely to have direct access to the original HDL codes. A hardware Trojan that leaks cryptographic keys has been inserted with the tool; nonetheless, it requires human effort for insertion, which hinders the production of a large HT dataset~\cite{cruz2022automatic}. \changes{Further endeavors have been made to follow a threat model in which an adversary is located in a foundry with sophisticated reverse-engineering capabilities. Perez~\etal~\cite{perez2021side} targets SCTs (Side-channel Trojans), more commonly found in crypto cores. The authors showcase a flow to insert HTs to leak confidential information based on power signatures. During this process, an adversary takes advantage of ECO (engineering change order), a flow originally used to fix bugs in finalized layouts. The work in \cite{perez2022hardware} builds upon the previous study by manufacturing an ASIC prototype with 4 HT-infected versions of AES and PRESENT. Puschner~\etal~\cite{puschner2023red} propose a de-coupled insertion and detection flow where the red team is responsible for inserting ECO-based HTs in design layouts, and the blue team must find the malicious embedding by investigating SEM (Scanning Electron Microscope) images vs GDSII (Graphic Design System II) files. The study shows that the ECO-inserted HTs are less challenging to find. Hepp~\etal~\cite{hepp2022pragmatic} use the ECO flow to insert HTs in the design layout without prior knowledge of its functionality. 
The study explores three new criteria for selecting the HT payload and triggers: transition probability, imprecise information flow tracking of selected signals, and the RELIC score. The RELIC score is a metric that provides an attacker with information about the location of a flip-flop relative to the data path or the control path. The authors operate under the assumption of a 24-hour time window for the attacker to complete the insertion process.}

Cruz~\etal~\cite{cruz2018automated} tried to address the benchmark shortcomings by presenting a toolset capable of inserting a variety of HTs based on the parameters passed to the toolset. Their software inserts HTs with the following configuration parameters: the number of trigger nets, the number of rare nets among the trigger nodes, a rare-net threshold (computed with functional simulation), the number of the HT instances to be inserted, the HT effect, the activation method, its type, and the choice of payload.  
Despite increasing the variety of inserted HTs, there is no solution for finding the optimal trigger and payload nets. The TRIT benchmark set generated by this tool is available on Trust-Hub~\cite{trusthub}. 

Cruz~\etal~\cite{cruz2022automatic} propose MIMIC, an ML framework for automatically generating Trojan benchmarks. The authors extracted $16$ functional and structural features from existing Trojan samples. Then, they trained ML models and generated a large number of hypothetical Trojans called \textit{virtual Trojans} for a given design. The virtual Trojans are then compared to a reference Trojan model and ranked. Finally, the selected Trojan will be inserted into the target circuit using suitable trigger and payload nets. The HT insertion process is highly convoluted, requiring multiple stages and expertise. MIMIC is not released publicly, and rebuilding the tool from their work is an extensive process. 
MIMIC's HT insertion criteria are very similar to~\cite{cruz2018automated}, and it suffers the same shortcomings~\cite{cruz2018automated}. 

To deceive machine learning HT detection approaches, Nozawa~\etal~\cite{nozawa2021generating} have devised adversarial examples. Their proposed method replaces the HT instance with its logically equivalent circuit, so the classification algorithm erroneously disregards it. To design the best adversarial example, the authors have defined two parameters: Trojan-net concealment degree (TCD), which is tuned to maximize the loss function of the neural network in the detection process, and a modification evaluating value (MEV) that should be minimized to have the least impact on circuits. These two metrics help the attacker to look for more effective logical equivalents and diversify HTs. The equivalent HTs are inserted in trust-hub benchmarks, and they decrease accuracy significantly.

\begin{table*}[!t]
\begin{center}
\scalebox{0.99}{
\begin{tabular}{|c|c|c|c|c|c|}
\hline
\bf{Tool} & \bf{Domain} & \bf{Insertion Criteria} & \bf{Automate} & \bf{open-source} \\ \hline\hline
Trust-Hub~\cite{shakya2017benchmarking}&  ASIC/FPGA & Secret Leakage, Signal Prob. & \ding{56} &  \ding{56}    \\ \hline
HAL~\cite{wallat2017look}&  ASIC/FPGA & Neighborhood Control Value & \ding{56} & \ding{52}    \\ \hline
TAINT~\cite{jyothi2017taint} & FPGA &Not Mentioned & \ding{56} & \ding{56} \\ \hline
TRIT~\cite{cruz2018automated}&  ASIC & Signal Prob. & \ding{56} & \ding{56}    \\ \hline
Yu~\etal~\cite{yu2019improved} & ASIC & Transition Prob. & \ding{52} & \ding{56} \\ \hline
Nozawa~\etal~\cite{nozawa2021generating} & ASIC & Same as~\cite{shakya2017benchmarking}& \ding{56} & \ding{56} \\ \hline
MIMIC~\cite{cruz2022automatic} & ASIC & Struct. \& Funct. Features & \ding{52} & \ding{56} \\ \hline
Sarihi~\etal~\cite{sarihi2022hardware} & ASIC & SCOAP paarameters & \ding{52} & \ding{56} \\ \hline
ATTRITION~\cite{gohil2022attrition} & ASIC & Signal Prob. & \ding{52} & \ding{56} \\ \hline

\textcolor{black}{Perez~\etal~\cite{perez2021side,perez2022hardware} }& \textcolor{black}{ASIC} & \textcolor{black}{Power Leakage} & \textcolor{black}{\ding{56}} & \textcolor{black}{\ding{56}} \\ \hline

\textcolor{black}{Puschner~\etal~\cite{puschner2023red} }& \textcolor{black}{ASIC} & \textcolor{black}{No restrictions} & \textcolor{black}{\ding{52}} & \textcolor{black}{\ding{52}} \\ \hline

\textcolor{black}{BioHT~\cite{hepp2022pragmatic} }& \textcolor{black}{ASIC} & \textcolor{black}{Multiple criteria} & \textcolor{black}{\ding{52}} & \textcolor{black}{\ding{56}} \\ \hline
\end{tabular}}
\caption{Survey of previous HT insertion tools.}
\label{tab_insertion}
\end{center}
\end{table*}

Sarihi~\etal~\cite{sarihi2022hardware} (our prior work) inserted a large number of HTs into ISCAS-85 benchmarks with Reinforcement Learning. The HT circuit is an agent that interacts with the environment (the circuit) by taking five different actions (next level, previous level, same level up, same level down, no action) for each trigger input. Level denotes the logic level in the combinational circuits. The agent moves the Trojan inputs throughout the circuit and explores various locations suitable for embedding HTs. Triggers are selected according to a set of SCOAP (Sandia Controllability/Observability Analysis Program~\cite{goldstein1980scoap}) parameters, \ie, a combination of controllability and observability. The agent is rewarded in proportion to the number of circuit inputs it can engage in the HT activation process. 

Gohil~\etal~\cite{gohil2022attrition} proposed ATTRITION, another RL-based HT insertion platform where signal probability is the target upon which the trigger nets are selected. The agent tries to find a set of so-called~\textit{compatible} rare nets, \ie, a group of rare nets that can be activated together with an input test vector. The test vector is generated using an SAT-solver. The authors also propose a pruning technique to limit the search space for the agent to produce more HTs in a shorter period. The tool is claimed to be open-source, but only the source code was released. 

Table~\ref{tab_insertion} summarizes the existing artifacts and research in the HT insertion space. It represents the target technology ($2^{nd}$ column); summarizes the insertion criteria ($3^{rd}$ column); shows if the tool is automated ($4^{th}$ column) and if the tool or its artifacts are openly released ($5^{th}$ column).

\subsection{Hardware Trojan Detection}
\label{subsec:detection_background}

Chakraborty~\etal~\cite{chakraborty2009mero} introduced MERO, a test vector generator that tries to trigger possible HTs by exciting rare-active nets multiple times. The algorithm's efficacy is tested against randomly generated HTs with rare triggers. MERO's detection rate significantly shrinks as circuit size grows. 

Hasegawa~\etal~\cite{hasegawa2017trojan} have proposed an ML method for HT detection. The method extracts $51$ circuit features from the trust-hub benchmarks to train a random forest classifier that eventually decides whether a design is HT-free. The HT classifier is trained on a limited HT dataset with an inherent bias during its insertion phase. 

Lyu~\etal~\cite{lyu2020scalable} proposed TARMAC to map the trigger activation problem to the clique cover problem, \ie, treating the netlist as a graph. They utilized an SAT-solver to generate the test vector for each maximal satisfiable clique. The method lacks scalability as it should run on each suspect circuit separately. Also, the achieved performance is not stable~\cite{pan2021automated}. Implementation of the method is neither trivial nor available publicly to researchers~\cite{gohil2022attrition}.

TGRL is an RL framework used to detect HTs~\cite{pan2021automated}. The agent decides to flip a bit in the test vector according to an observed probability distribution. The reward function, which combines the number of activated nets and their SCOAP~\cite{goldstein1980scoap} parameters, pushes the agent to activate as many signals as possible. 
Despite its higher HT detection rate than MERO and TARMAC, the algorithm was not tested on any HT benchmarks~\cite{gohil2022attrition}.

DETERRENT, an RL-based detection method~\cite{gohil2022deterrent}, finds the smallest set of test vectors to activate multiple combinations of trigger nets. The RL state is a subset of all possible rare nets, and actions are appending other rare nets to this subset. The authors used an SAT-solver to determine if actions are compatible with the rare nets in the subsets, and they only focused on signal-switching activities as their target. 

The HW2VEC tool~\cite{yu2021hw2vec} converts RTL-level and gate-level designs into a dataflow graph and abstract syntax tree to extract a feature set that represents the structural information of the design. Extracted features are used to train a graph neural network to determine whether a design is infected with HTs. The authors test the tool with $34$ circuits infected by in-house generated HTs.  

We note that of the methods reviewed above (and others studied but not discussed here), the only publicly available tool is HW2VEC.
Table~\ref{tab_detection} summarizes the previous works in HT detection where researchers have used various criteria in detecting HTs ($2^{nd}$ column) and the open-source state of the work ($3^{rd}$ column).  

\begin{table}[]
\caption{Survey of Previous HT Detection Tools.}
\label{tab_detection}
\begin{tabular}{|c|c|c|}
\hline
\bf{Study} & \bf{Detection Basis} & \bf{open-source}  \\ \hline\hline
MERO~\cite{chakraborty2009mero}&  Switching Activity  & \ding{56}            \\ \hline
Hasegawa~\etal~\cite{hasegawa2017trojan} &    Netlist Features    &     \ding{56}  \\     \hline
TARMAC~\etal~\cite{lyu2020scalable}  &    Switching Activity    &     \ding{56}            \\ \hline
TGRL~\etal~\cite{pan2021automated} &     Switching Activity     &     \ding{56}             \\ \hline
DETERRENT~\etal~\cite{gohil2022deterrent}  &    Switching Activity    &     \ding{56}             \\ \hline
HW2VEC~\cite{yu2021hw2vec} & Graph Structural Info. & \ding{52} \\ \hline
\end{tabular}
\end{table}\section{The Proposed HT Insertion}
\label{sec:insertion}

\begin{figure*}[!t]
  \centering
  \includegraphics[width=\textwidth]{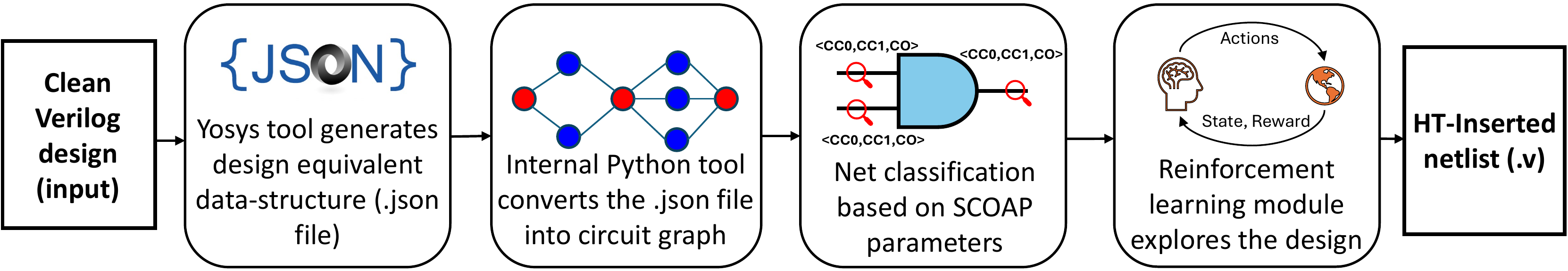}
  \caption{The proposed RL-based HT insertion tool flow.}
  \vspace{-5mm}
  \label{fig:Insertion_flow}
\end{figure*}

Figure~\ref{fig:Insertion_flow} shows the flow of the proposed HT insertion tool. The first step creates a graph representation of the flattened netlist from the circuit. Yosys Open Synthesis Suite~\cite{wolf2013yosys} translates the HDL (Verilog) source of the circuit into a JSON (JavaScript Object Notation)~\cite{bassett2015introduction} netlist, which enables us to parse the internal graph representation of the circuit. Next, the tool finds a set of rare nets to be used as HT trigger nets (this step is described in detail in Subsection~\ref{subsec:rare_net_extraction}). Finally, an RL agent uses the rare net information and attempts to insert an HT to maximize a rewarding function as described in Section~\ref{subsec:RL_HT_insertion}.

\subsection{Rare Nets Extraction}
\label{subsec:rare_net_extraction}

We use the parameters introduced in~\cite{sebt2018circuit} to identify trigger nets. These parameters are defined as functions of net \emph{controllability} and \emph{observability}. Controllability measures the difficulty of setting a particular net in a design to either \emph{'0'} or \emph{'1'}. Conversely, observability is the difficulty of propagating a net value to at least one of the circuit's primary outputs~\cite{goldstein1980scoap}. 

The first parameter is called the HT trigger susceptibility parameter, and it is derived from the fact that low-switching nets have mainly a high difference between their controllability values. Equation~\ref{HTS1} describes this parameter:
\begin{equation}
    HTS(Net_i)=\frac{|CC1(Net_i)-CC0(Net_i)|}{Max(CC1(Net_i),CC0(Net_i))}
    \label{HTS1}
\end{equation}
where $HTS$ is the HT trigger susceptibility parameter of the net; $CC0(Net_i)$ and $CC1(Net_i)$ are the combinational controllability $0$ and $1$ of $Net_i$, respectively. The $HTS$ parameter ranges between $[0,1)$ such that higher values correlate with lower activity on the net. 

The other parameter, specified in Equation~\ref{OCR}, measures the ratio of observability to controllability:
\begin{equation}
    OCR(Net_i)=\frac{CO(Net_i)}{CC1(Net_i)+CC0(Net_i)}
    \label{OCR}
\end{equation}
where $OCR$ is the observability to controllability ratio. This equation requires that the HT trigger nets be hard to control but not so hard to observe. Unlike the $HTS$ parameter, $OCR$ is not bounded and belongs to the $[0,\infty)$ interval. We will specify thresholds (see Section~\ref{sec:Result}) for each parameter and use them as filters to populate the set of rarely-activated nets for our tool.

\subsection{RL-Based HT Insertion}
\label{subsec:RL_HT_insertion}

\begin{figure}[!t]
  \centering
  \includegraphics[width=.49\textwidth]{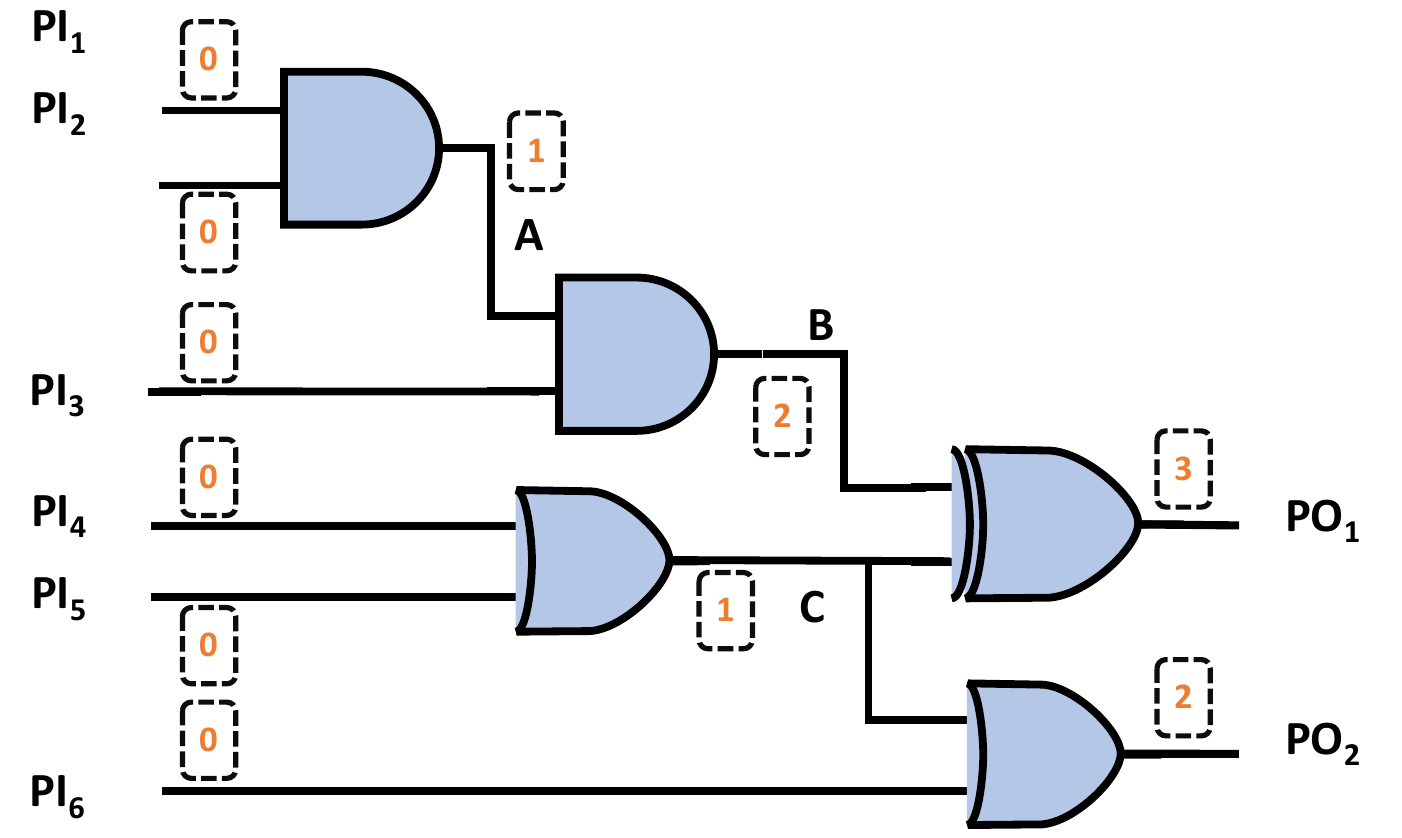}
  \caption{Levelizing a circuit. The output level of each digital gate is computed by $max(Level(in1),Level(in2))+1$.}
  \label{level}
\end{figure}

The RL environment is, in fact, the circuit in which the agent is trying to insert HTs. The agent's action is to insert combinational HTs where trigger nets are ANDED, and the payload is an XOR gate (same as Figure~\ref{trigger_payload}). The RL agent starts from a reset condition, taking a series of actions that eventually insert HTs in the circuit. Different HT insertion options are represented with a state vector in each circuit. For a given HT, the state vector is comprised of $s_t=[s_1,s_2, ...,s_{n-2},s_{n-1},s_{n}]$ where $s_1$ through $s_{n-2}$ are the logic-levels of the HT inputs, and $s_{n-1}$ and $s_{n}$ are the logic-levels of the target net and the output of the XOR payload, respectively. Figure~\ref{level} shows how we conduct the circuit levelization. Here, the circuit Primary Inputs (PIs) are considered level $0$. The output level of each gate is computed by Equation~\ref{eq:level}:
\begin{equation}
    Level(output)=MAX(Level(in_1), Level(in_2))+1
    \label{eq:level}
\end{equation}

\begin{figure}[!t]
  \centering
  \includegraphics[width=.49\textwidth]{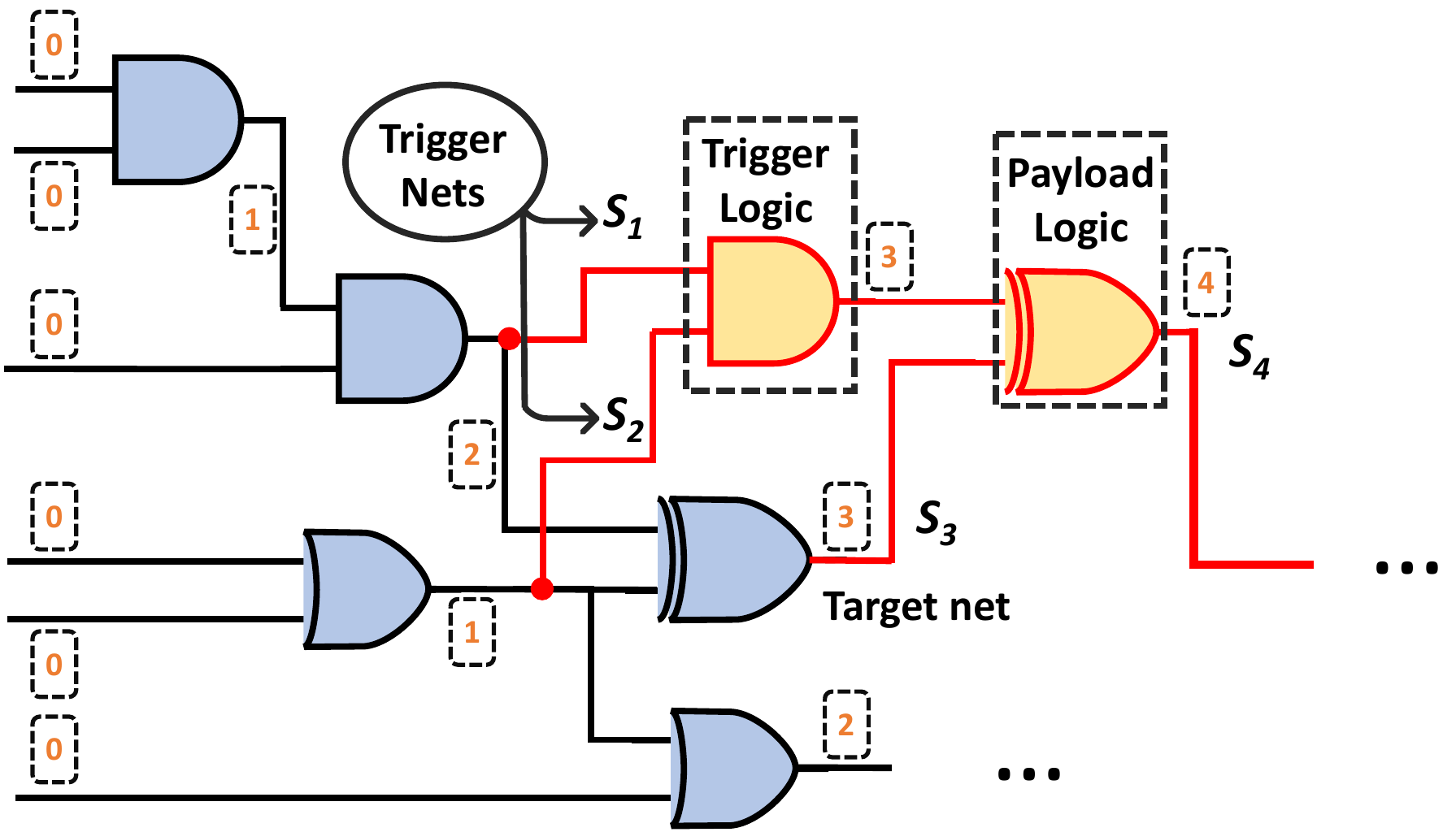}
  \caption{Obtaining the state vector in the presence of an HT in the circuit.}
  \label{State}
\end{figure}

As an example, the HT in Figure~\ref{State} (in yellow) has the state vector $s_t=[2,1,3,4]$. The action space of the described HT agent is multi-discrete, \ie, each input of the HT may choose an action from a set of five available actions. These actions are:

\begin{itemize}
    \item \emph{\textbf{Next level}}: the input of the HT moves to one of the nets that are one level higher than the current net level.
    \item \emph{\textbf{Previous level}}: the input of the HT moves to one of the nets that are one level lower than the current net level.
    \item \emph{\textbf{Same level up}}: the input of the HT will move to one of the nets at the same level as the current net level. The net is picked by pointing to the next net in the ascending list of net IDs for the given level. 
    \item \emph{\textbf{Same level down}}: the input of the HT will move to one of the nets at the same level as the current net level. The net is picked by pointing to the previous net in the ascending list of nets for the given level. 
    \item \emph{\textbf{No action}}: the input of the HT will not move. If an action leads the agent to step outside the circuit boundaries, it is substituted with a ``No action''.
\end{itemize}

\begin{algorithm}[t]
    \caption{Training of the HT inserting Reinforcement Learning Agent}
    \begin{flushleft}
    \hspace*{\algorithmicindent}\textbf{\textit{Input: }}{Graph $G$, HTS Threshold $T_{HTS}$, OCR Threshold}\\
    \hspace*{\algorithmicindent}{$T_{OCR}$, Circuit Inputs $in\_ports$, State Space $s_t$,}\\
    \hspace*{\algorithmicindent}{Terminal State $Terminal_{state}$, Total Timesteps $j$;}\\
    \hspace*{\algorithmicindent}\textbf{\textit{Output: }}{HT Benchmark $HT_{Benchmark}$;}
    \end{flushleft}
         
     \begin{algorithmic}[1]
     \STATE Compute SCOAP parameters:\\
     \hspace*{\algorithmicindent}{$<CC0,CC1,CO>=computeSCOAP(G)$};
     \STATE Get the set of rare nets:\\
     \hspace*{\algorithmicindent}{$rare\_nets=Compute\_Rare\_Nets(G, T_{HTS}, T_{OCR});$}\\
     \STATE $counter=0;$
     \WHILE{($counter<j$)}
        \STATE $HT=reset\_environment();$
        \STATE $Terminal_{state}=false;$
        \WHILE{!($Terminal_{state}$)}
            \STATE $G, s_t, Terminal_{state}, HT_{triggers}=action(HT);$
            \STATE $HT\_activated =PODEM(G);$
                \STATE $temp_{reward} = (HT_{triggers} \cap rare\_nets).count();$
                \IF {$(HT\_activated)$}
                    \IF {$(temp_{reward}==1)$}
                        \STATE $reward=8;$
                    \ELSIF{$(temp_{reward}==2)$}
                        \STATE $reward=16;$
                    \ELSIF{$(temp_{reward}==3)$}
                        \STATE $reward=100;$
                    \ELSIF{$(temp_{reward}==4)$}
                        \STATE $reward=1000;$
                    \ELSIF{$(temp_{reward}==5)$}
                        \STATE $reward=10000;$
                    \ELSE 
                        \STATE $reward=-1;$
                    \ENDIF
                \ENDIF
            \STATE $update\_PPO(action,s_t,reward);$
            \STATE $counter+=1;$
        \ENDWHILE
        
     \ENDWHILE
     \STATE $HT_{Benchmark}=Graph\_to\_netlist(G)$
    \end{algorithmic}
    \label{alg:HT_insertion}    
\end{algorithm}

The action space is also represented by a vector where its size is equal to the number of the HT inputs, and each action can be one of the five actions above, \eg, for the HT in Figure~\ref{State}, the action space would be $a_t=[a_1,a_2]$ since it has two inputs. Hypothetical actions for the first and the second inputs can be the same level up/down and next/previous level, respectively. 

The flow of our RL inserting agent is described in Algorithm~\ref{alg:HT_insertion}. The SCOAP parameters are first computed (line $1$). We specify two thresholds $T_{HTS}$ and $T_{OCR}$ and require our algorithm to find nets that have higher $HTS$ values than $T_{HTS}$ and lower $OCR$ values than $T_{OCR}$ (line $2$). These nets are classified as rare nets. The algorithm consists of two nested while loops that keep track of the terminal states and the elapsed timesteps. The latter defines the total number of samples the agent trains on. We have used the OpenAI Gym~\cite{DBLP:journals/corr/BrockmanCPSSTZ16} environment to implement our RL agent. 

The first used method is called $reset\_environment()$, which resets the environment before each episode and returns the initial location of the agent HT (line $5$). The HT is randomly inserted within the circuit according to the following rules.

\begin{itemize}
    \item Rule 1) Trigger nets are selected randomly from the list of the total nets.
    \item Rule 2) Each net can drive a maximum of one trigger net.
    \item Rule 3) Trigger nets cannot be assigned as the target.
    \item Rule 4) The target net is selected with respect to the level of trigger nets. To prevent forming combinational loops, we specify that the level of the target net should be greater than that of the trigger nets. 
\end{itemize}

\begin{figure*}[!t]
  \centering
  \includegraphics[width=\textwidth]{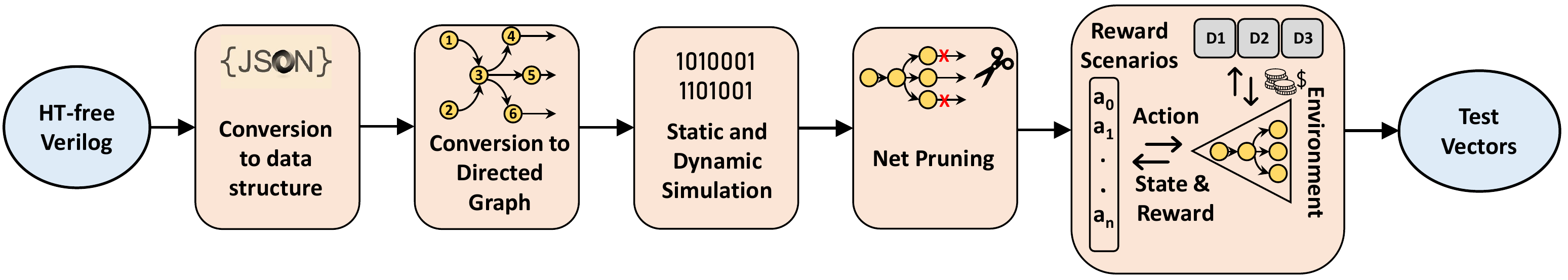}
  \caption{The proposed detection flow.}
  \vspace{-6mm}
  \label{flow}
\end{figure*}

In each episode of the training process, we keep the target net unchanged to help the RL algorithm converge faster. Instead of manually specifying a target net, we let the algorithm explore the environment and choose a target net. 
The terminal state variable $TS$ is set to $False$ to check the termination condition for each episode. When the trigger nets' level reaches the target net's level, or the number of steps per episode reaches an allowed maximum (lines $6$-$7$), $TS$ becomes $True$, which terminates the episode.

The training process of the agent takes place in a loop where actions are being issued, rewards are collected, the state is updated, and eventually, the updated graph is returned. To test the value of an action taken by the RL agent (meaning if the HT can be triggered with at least one input pattern), we use $PODEM$ (Path-Oriented Decision Making), an automatic test pattern generator~\cite{bushnell2000essentials} (line $9$). This algorithm uses a series of backtracing and forward implications to find a vector that activates the inserted HT. If the HT payload propagates through at least one of the circuit outputs, the action gains a reward proportional to the number of rare triggers on the HT. After the number of rare triggers is counted in line $10$, the agent is rewarded in lines $11$ through $25$. The rewarding scheme is designed such that the agent would start finding HTs with a $1$ rare trigger net and adding more rare nets while exploring the environment.
Additionally, the exponential reward increase in each case ensures that the agent is highly encouraged to find HTs with at least three or more rare trigger nets. If an HT is not activated with $PODEM$ or no rare nets are among the HT triggers, the agent will be rewarded $-1$. Since the agent is unlikely to find high-reward HTs at the beginning of the exploration stage, the first two rewarding cases ($temp_{reward}=1$ and $temp_{reward}=2$) should be set such that the agent sees enough positive, rewarding improvements, yet be more eager to find more HTs that yield higher rewards. After extensive experiments with the RL agent, the reward values are assigned to different cases.

We use the $PPO$ (Proximal Policy Optimization)~\cite{schulman2017proximal} RL algorithm to train the RL agent. PPO can train agents with multi-discrete action spaces in discrete or continuous spaces. The main idea of PPO is that the new updated policy (which is a set of actions to reach the goal) should not deviate too far from the old policy following an update in the algorithm. To avoid substantial updates, the algorithm uses a technique called clipping in the objective function~\cite{schulman2017proximal}. Using a clipped objective function, PPO restricts the size of policy updates to prevent them from deviating too much from the previous policy. This constraint promotes stability and ensures that the updates are controlled within a specific range, which helps avoid abrupt changes that may negatively affect the agent's performance. At last, when the HTs are inserted, the toolset outputs Verilog gate-level netlist files that contain the malicious HTs (line $30$).

\begin{figure}[!b]
  \centering
  \includegraphics[width=.49\textwidth]{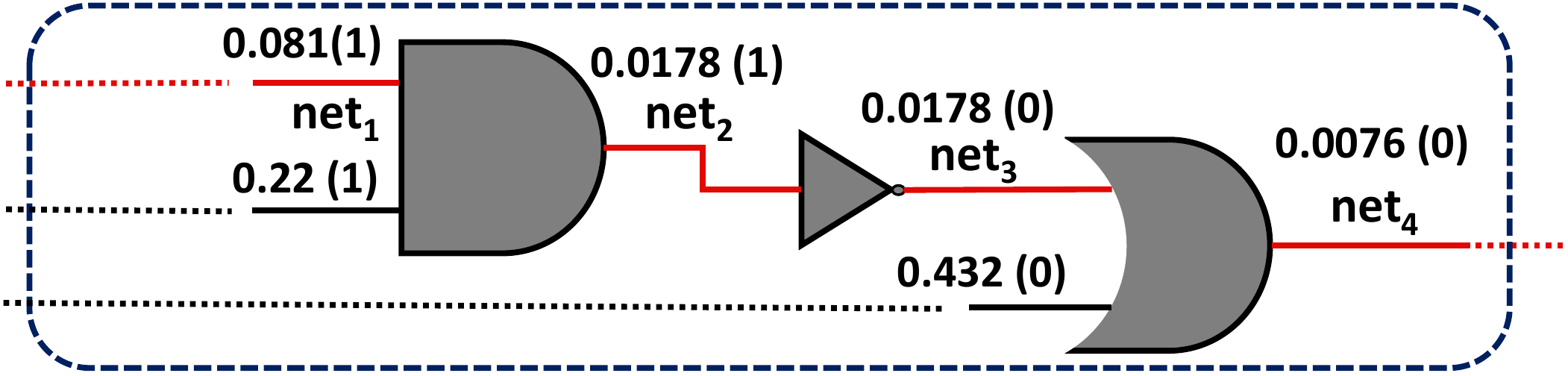}
  \caption{State pruning identifies nets in the same activation path.}
  \label{Pruned}
\end{figure}

\section{The Proposed HT Detection}
\label{sec:detection}

From a detection perspective, we must determine whether a given circuit is clean or Trojan-infected. To achieve this goal, an RL agent is defined that applies its generated test vectors to circuits and checks for any deviation at the circuits' primary outputs with respect to the expected outputs (golden model). The agent interacts with the circuit (performs actions) by flipping the vector values to activate certain internal nets. The action space is an $n$-dimensional binary array where $n$ is the number of circuit primary inputs. The action space vector $a_t$ is defined as $a_t=[a_1,a_2,..., a_n]$. The agent decides to toggle each $a_i$ to transition to another state or leave them unchanged. $a_i=0$ denotes that the value of the $i^{th}$ bit of the input vector should remain unchanged from the previous test vector.
In contrast, $a_i=1$ means that the $i^{th}$ input bit should flip. The RL agent follows a $\pi$ policy to decide which actions should be commenced at each state. The $\pi$ policy is updated using a policy gradient method~\cite{nguyen2019deep} where the agent commences actions based on probability distribution from the $\pi$ policy. The assumption is that attackers are likely to choose trigger nets with a consistent value ($0$ or $1$) most of the time. Thus, a detector aims to activate as many dormant nets as possible. We consider two different approaches for identifying such rare nets:

\textbf{1) Dynamic Simulation}: We feed each circuit with $100$K random test vectors and record the value of each net. Then, we populate the switching activity statistics during the simulation time and set a threshold $\theta$ for rare nets where the switching activity for a net below $\theta$ denotes that the net is rare. $\theta$ is in the range of $[0,1]$.

\textbf{2) Static Simulation}: We use the $HTS$ parameter in Equation~\ref{HTS1} and a threshold to find rare nets. Categorizing rare nets with this approach provides the security engineer with an extra option for detection.

In a circuit with $m$ rare nets, the state space is defined as $State_t=[s_1, s_2, ..., s_m]$ where $s_i$ is associated with the $i^{th}$ net in the set. If an action (a test vector) sets the $i^{th}$ net to its rare value, $s_i$ will be $1$; otherwise, $s_i$ stays at $0$. As can be inferred, the action and state spaces are multi-binary. 

Attackers tend to design multi-trigger HTs~\cite{cruz2018automated}, and this should be considered when HT detectors are designed. The final purpose of our detector is to generate a set of test vectors that can trigger as many rare nets as possible. To achieve this goal, a part of the rewarding function should enumerate rare nets. However, we should avoid over-counting situations where a rare net has successive dependent rare nets. An example case is shown in Figure~\ref{Pruned} where four nets $net_{1}$, $net_{2}$, $net_{3}$, and $net_{4}$ (with their switching probabilities and their rare values) are all dependent rare nets. Instead of including all four nets in the state space, we choose the rarest net as the representative net since activating the rarest net ensures the activation of the others as well. In this example, $net_{4}$ is selected as the set representative. This policy helps accelerate the RL agent to converge on the global minima faster. Figure~\ref{flow} summarizes our proposed detection flow. 

As for rewarding the agent, we consider three rewarding functions, which we explain here. Our multi-rewarding detector enables security engineers to better prepare for attackers with different mindsets.

\subsection{Rewarding function SSD}
\begin{figure}[!t]
  \centering
  \includegraphics[width=0.49\textwidth]{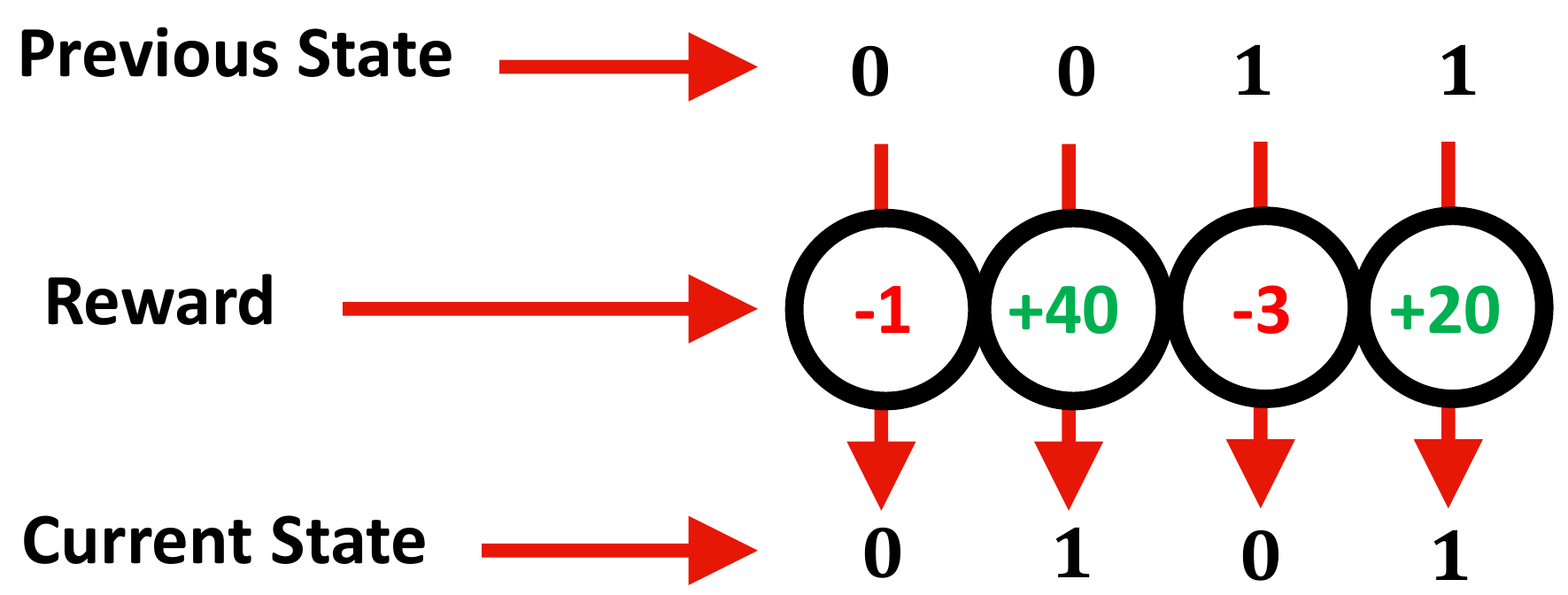}
  \caption{Rare net transition (state transition) in the current and previous states and corresponding rewards}
  \label{fig:D1}
\end{figure}

In our first rewarding function (Algorithm~\ref{alg:D1}) called SSD (Subsequent State Detector), we push the RL agent to build on its current state. We use a copy of the previous state and encourage the agent to generate state vectors that differ from the previous one. The hypothesis is to push the agent toward finding test vectors that lead to various unseen states. 
To compute the reward, the pruned current and previous state vectors and their lengths are passed as inputs to Algorithm~\ref{alg:D1}. The rewarding function comprises an $immediate$ and a $sequential$ part, initialized to $0$ in lines $1$ and $2$, respectively. Whenever the state transitions, we iterate through the loop $K$ times. We calculate the sequential reward by making a one-to-one comparison between the nets in the old and new states. In lines $5-11$, the highest reward is given when an action can trigger a net not triggered in the previous state, \ie, $+40$. If a rare net is still activated in the current state, the agent will still get rewarded $+20$. The worst state transition is whenever an action leads to a rare net losing its rare value, which is rewarded $-3$. Lastly, if the agent cannot activate a rare net after a state transition, it will be rewarded $-1$. This process is depicted in Figure~\ref{fig:D1}.
The immediate award is the number of activated rare nets in the new state. The ultimate reward value is a linear combination of the immediate and sequential rewards with coefficients $\lambda_{1}$ and $\lambda_{2}$, respectively, which are tunable parameters to be set by the user. We build the state vector with the obtained rare nets from functional simulation.

\begin{algorithm}[t]
    \caption{Rewarding Function SSD}
    \begin{flushleft}
    \hspace*{\algorithmicindent}\textbf{\textit{Input: }}{$State_{pre}$, $State_{cur}$, State Vector Length $K$}\\
    \hspace*{\algorithmicindent}\textbf{\textit{Output: }}{$Reward_{final}$}\\
    \end{flushleft}
     \begin{algorithmic}[1]
        \STATE $Reward_{Imd} = 0;$
        \STATE $Reward_{Seq} = 0;$
        \FOR{$k \in \{0,\dots,K-1\}$}
            \IF {($State_{cur}[k]=0$ and $State_{pre}[k]=0$)}
                \STATE $Reward_{Seq} += -1;$
            \ELSIF{($State_{cur}[k]=0$ and $State_{pre}[k]=1$)}
                \STATE $Reward_{Seq} += -3;$
            \ELSIF{($State_{cur}[k]=1$ and $State_{pre}[k]=0$)}
                \STATE $Reward_{Seq} += 40;$
            \ELSIF{($State_{cur}[k]=1$ and $State_{pre}[k]=1$)}
                \STATE $Reward_{Seq} += 20;$
            \ENDIF
        \ENDFOR
        \STATE $Reward_{Imd.} = State_{cur}.count(1)  $
        \STATE $Reward_{final}=\lambda_{1}\times Reward_{Seq} +\lambda_{2}\times Reward_{Imd}$
    \end{algorithmic} 
    \label{alg:D1}
\end{algorithm}

\subsection{Rewarding function SAD}
Algorithm~\ref{alg:D2} describes our second rewarding function called SAD (Switching Activity Detector). In this case, the agent gains rewards proportional to the difficulty of the rare nets triggered. First, the reward vector is initiated with a length equal to the state vector (line $1$). Each element in the reward vector has a one-to-one correspondence with rare nets on the state vector. The reward for each rare net is computed by taking the inverse of the net switching activity rate (line $4$).
In some cases, a net might have a switching probability of $0$. In such cases, activating the net would be rewarded 10X times the greatest reward in the vector (line $12$). Thus, upon observing every new state, the agent will be rewarded based on the activated nets and the reward vector (line $18$). If a rare net is not activated, -1 will be added to the final reward (line $20$). The algorithm encourages the agent to trigger the rarest nets in the circuit directly.

\begin{algorithm}[!t]
    \caption{Rewarding Function SAD}
    \begin{flushleft}
    \hspace*{\algorithmicindent}\textbf{\textit{Input: }}{Net switching vector} $Switching_{vector}$,\\ \hspace*{\algorithmicindent}\textbf{}{Current state vector $State_{vector}$, State Vector Length $K$ }\\
    \hspace*{\algorithmicindent}\textbf{\textit{Output: }}{Final reward $Reward_{final}$}\\
    \end{flushleft}
     \begin{algorithmic}[1]
        \STATE $Reward_{vector} = [0] * K$ 
        \FOR{$k \in \{0,\dots,K-1\}$}
            \IF {($Switching_{vector}[k] != 0$)}
                \STATE $Reward_{vector}[k] = Switching_{vector}[k]^ {-1} $
            \ELSE
                \STATE $Reward_{vector}[k]= 0$
            \ENDIF
        \ENDFOR
        \STATE $reward_{max} = max(Reward_{vector}[~]$)
        \FOR{$k \in \{0,\dots,K-1\}$}
            \IF {($Switching_{vector}[k] == 0$)}
                \STATE $Reward_{vector}[k] = 10 * reward_{max}  $
            \ENDIF
        \ENDFOR
        \STATE $Reward_{final} = 0$
        \FOR{$k \in \{0,\dots,K-1\}$}
            \IF {($State_{vector}[k]==1$)}
                \STATE $Reward_{final} += Reward_{vector}[k]$
            \ELSE
                \STATE  $Reward_{final} += -1$
            \ENDIF
        \ENDFOR
        
    \end{algorithmic}
    \label{alg:D2}    
\end{algorithm}

\subsection{Rewarding function COD}
\begin{figure}
    \centering
\includegraphics[width=.49\textwidth]{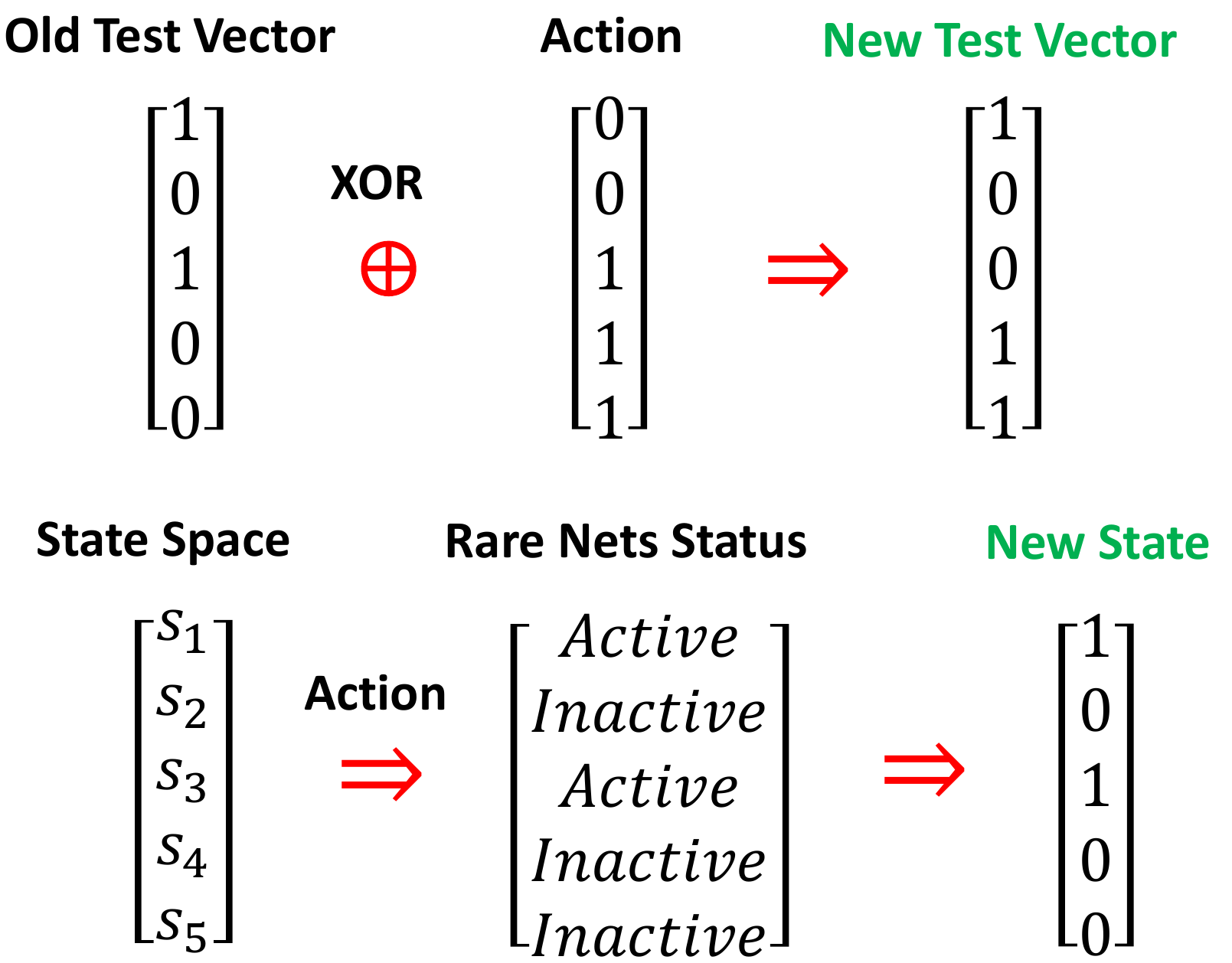}
    \caption{Test vector generation and state transition for SAD and COD}
    \label{fig:state_transition}
\end{figure}

The third rewarding function is described in Algorithm~\ref{alg:D3} and is called COD (Controllability Observability Detector). In this scenario, rare nets are populated based on the threshold of the $HTS$ parameter computed during the static simulation using Equation~\ref{HTS1}. When a rare net in the set is activated, the agent is rewarded with the controllability of the rare value (line $4$). Otherwise, it will receive $-1$ from the environment (line $6$). This scenario aims to investigate controllability-based HT detection with the RL agent. Figure~\ref{fig:state_transition} shows an example where an RL action is XORed with an old test vector, generating a new test vector. It also shows how activating rare nets (from SAD and COD) leads to state transitions where an activated net corresponds to a '1' in the state vector.

\begin{algorithm}[!t]
    \caption{Rewarding Function COD}
    \begin{flushleft}
    \hspace*{\algorithmicindent}\textbf{\textit{Input: }}{Controllability reward vector $Reward_{vector}$,}\\
    \hspace*{\algorithmicindent}\textbf{}{Current state vector $State_{vector}$,  State Vector Length $K$,   }\\
    \hspace*{\algorithmicindent}\textbf{\textit{Output: }}{Final reward $Reward_{final}$}\\
    \end{flushleft}
     \begin{algorithmic}[1]
        \STATE $Reward_{final} = 0$
        \FOR{$k \in \{0,\dots,K-1\}$}
            \IF {$State_{vector}[k]==1$}
                \STATE $Reward_{final} += Reward_{vector}[k]$
            \ELSE
                \STATE  $Reward_{final} += -1$
            \ENDIF
        \ENDFOR
        
    \end{algorithmic}
    \label{alg:D3}    
\end{algorithm}

\section{The Proposed Generic HT-Detection Metric}
\label{sec:detection_metric}

We propose the following methodology to the community for fair and repeatable comparisons among HT detection methods. In addition, our methodology can help compare different HT insertion techniques for a given HT detector. This methodology obtains a confidence value that one can use to compare different HT detection methods.  

\begin{figure}
    \centering
\includegraphics[width=.49\textwidth]{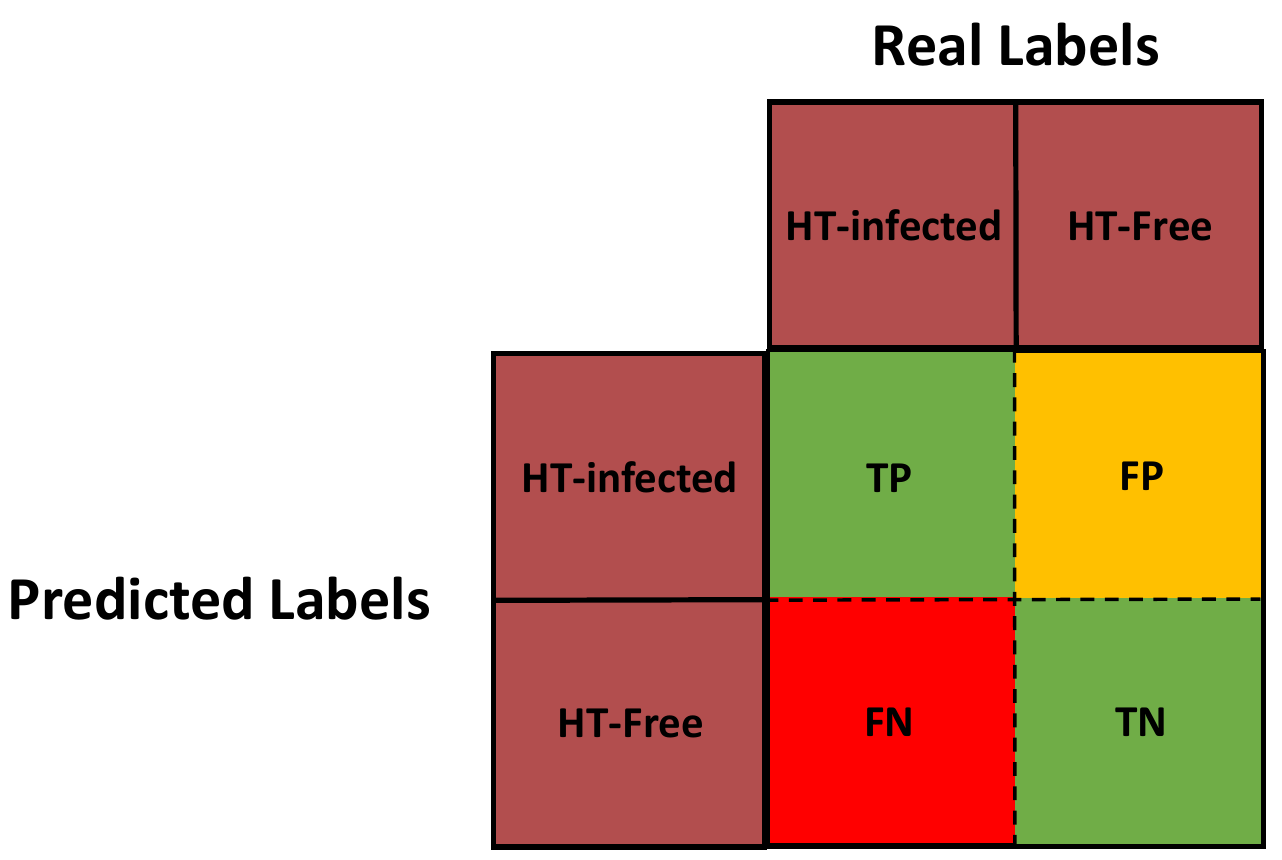}
    \caption{Possible outcomes of an HT detection trial.}
    \label{fig:TFPN}
\end{figure}

Figure~\ref{fig:TFPN} shows four possible outcomes when an HT detection tool studies a given circuit. From the tool user's perspective, the outcomes are probabilistic events. For example, when an HT-free circuit is being tested, the detection tool may either classify it as an infected or a clean circuit, \ie, $Prob(FP) + Prob(TN) = 1$ where $FP$ and $TN$ stand for \textit{False Positive} and \textit{True Negative} events. Similarly, for HT-infected circuits, we have $Prob(FN) + Prob(TP) =1$. $FN$ and $FP$ are two undesirable outcomes at which detectors misclassify the given circuit. However, the $FN$ cases pose a significantly greater danger as they result in a scenario where we rely on an HT-infected chip. In contrast, an $FP$ case means wasting a clean chip by either not selling or not using it. So, we need to know how the user of HT detection tools (might be a security engineer or a company representative) prioritizes $FN$ and $FP$ cases. We define a parameter $\alpha$ as the ratio of the undesirability of $FN$ over $FP$. The tool user determines $\alpha$ based on characteristics and details of the application that eventually chips will be employed in, \eg, the risks of using an infected chip in a device with a sensitive application versus using a chip for home appliances. Note that the user sets this value, which is not derived from the actual $FP$ and $FN$. After $\alpha$ is set, it is plugged into Equation~\ref{equ:metric} and a general confidence basis $ Conf. \, Val $ is computed. 

\begin{equation}
    \label{equ:metric}
    Conf. \, Val = \frac{(1-FP)}{(1/\alpha+FN)}
\end{equation}

\begin{figure}[!t]
  \centering
  \includegraphics[scale=.58]{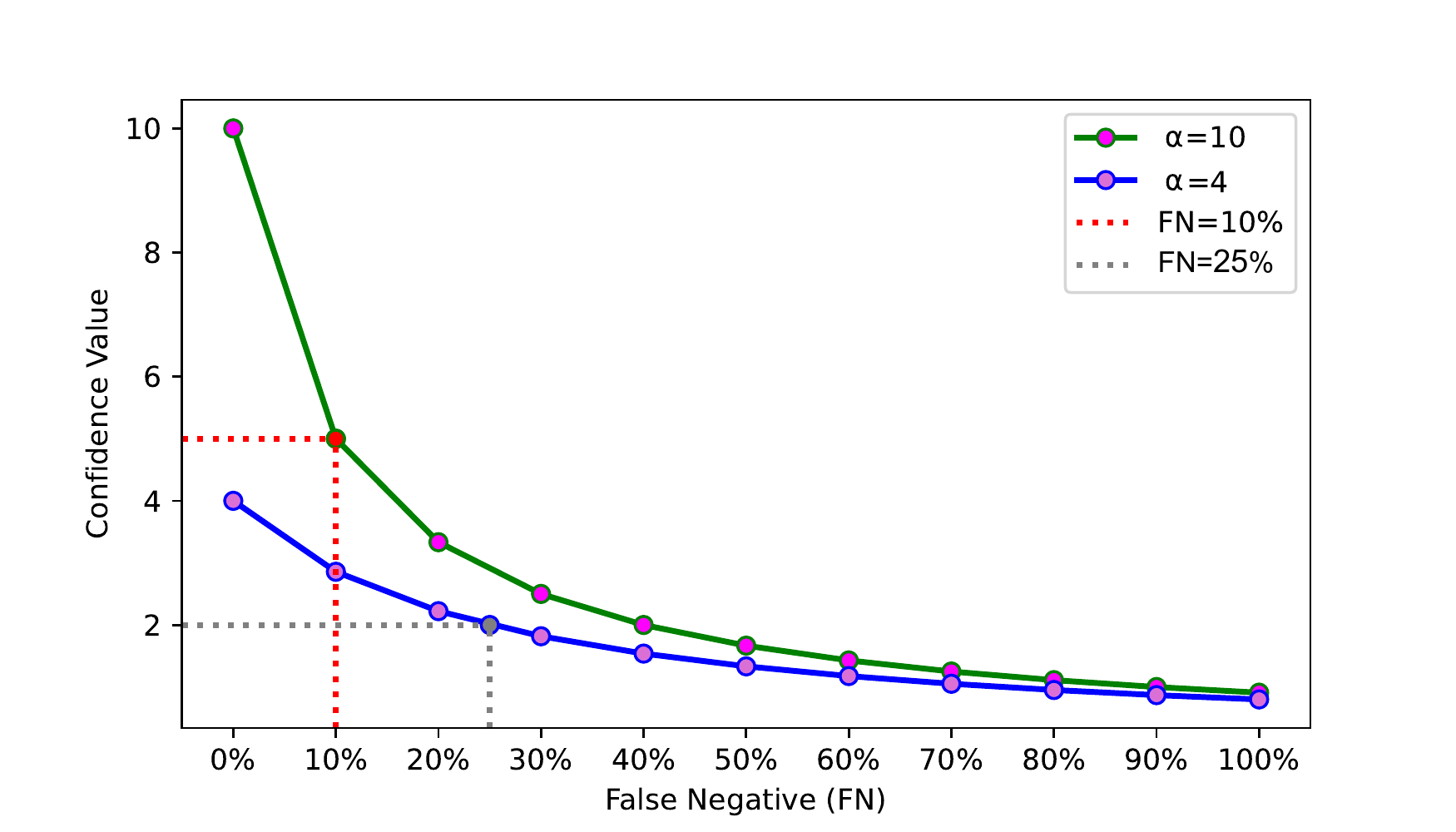}
  \caption{Confidence value vs.\ the percentage of \textit{FN} in our detectors assuming \textit{$\alpha=10$} and \textit{$\alpha=4$} }
  \vspace{-6mm}
  \label{fig:confidence}
\end{figure}

\begin{table*}[!t]
\caption{Characteristics of different circuits from ISCAS-85 benchmark}
\begin{center}
\scalebox{0.99}{
\begin{tabular}{|c|c|c|c|c|c|c|c|}
\hline
\textbf{Benchmark} & \textbf{\# of Inputs} & \textbf{\# of Levels} & \textbf{\# of nodes} & \textbf{\# of nets} & \textbf{$T_{OCR}$} & \textbf{$T_{HTS}$}  & \textbf{Description} \\ \hline 
c432  &   36  & 40 &  352    &  492  & 14 & 0.85  &  27-Channel Interrupt Controller   \\ \hline
c880  &   60  & 43 &  607    &  889  & 15 & 0.82  &  8-Bit ALU    \\ \hline
c1355 &   41  & 44 &  957    &  1416  & 20 & 0.75  &  32-Bit SEC Circuit  \\ \hline
c1908 &   33  & 52 &  868    &  1304  & 14 & 0.90  &  16-bit SEC/DED Circuit   \\ \hline
c2670 &   233  & 28 &  1323   &  1807  & 20 & 0.83  &  12-bit ALU and Controller   \\ \hline
c3540 &   50  & 60 &  1539   &  2527  & 15 & 0.84  &  8-bit ALU \\ \hline
c5315 &   178  & 63 &  2697   &  4292  & 21 & 0.79  &  9-bit ALU    \\ \hline
c6288 &    32 & 240 &  4496   &  6801  & 18 & 0.8 &  16x16 Multiplier \\  \hline
c7552 &    207  & 53 &  3561   &  5433  & 20 & 0.8  &32-Bit Adder/Comparator  \\ \hline
\end{tabular}
}
\end{center}
\label{tab:circuits}
\end{table*}

This metric can compare HT detection methods fairly regardless of their detection criteria and implementation methodology. The defined confidence metric combines the two undesirable cases to their severity from a security engineer's point of view. The $ Conf. \, Val $ ranges between $[\frac{0.5\alpha}{1+0.5\alpha}..\alpha]$. The closer the value is to $\alpha$, the more confidence in the detector. The absolute minimum of the $Conf. \, Val = 1/3$ that happens when $\alpha=1$ and $FP = FN = 50\%$ . This analysis assumes that $FN$ and $FP$ are independent probabilities. We note that, for some detection methods, $FP$ is always $0$. For instance, test-based HT detection methods that apply a test vector to excite HTs use a golden model (HT-free) circuit for comparison and decision-making, and it is impossible for a non-infected circuit to have a mismatch with the golden model (from the perspective of functional simulation). It is impossible for such methods to detect an HT in a clean circuit falsely. However, our metric is general and captures such cases.

Figure~\ref{fig:confidence} shows the relation between the confidence value and the $FN$ percentage for $\alpha=10$ and $\alpha=4$ for a test-based detector. As can be observed, the slopes of the graphs are different when $FN$ approaches zero. The maximum tolerable $FN$ is an upper bound for the $FN$ value at which we gain at least half the maximum confidence. As shown with the dashed lines in Figure~\ref{fig:confidence}, the maximum tolerable $FN$ for $\alpha=4$ and $\alpha=10$ is, respectively, $FN = 25\% $  and $FN = 10\% $.
Based on the figure, it can be inferred that choosing a higher base $\alpha$ will make it more challenging to attain higher confidence values. This fact should be considered when selecting $\alpha$ and interpreting the confidence values.

In addition to the detection quality, which the proposed confidence value can measure, HT detection methods should also be compared from a computational cost point of view. In particular, we encourage researchers to report the run-time of their methods and the training time, if applicable.

\section{Experimental Results and Discussion}
\label{sec:Result}

This section demonstrates the efficiency of the developed HT insertion and detection framework. For our experiments, we use an AMD EPYC 7702P 64-Core CPU with 512GB of RAM to train and test our agents. The training of the RL agents is done using the Stable Baselines library~\cite{raffin2019stable} with MLP (multi-layer perceptron) as the PPO algorithm policy~\cite{schulman2017proximal}. The benchmark circuits are selected from ISCAS-85~\cite{bryan1985iscas} and converted into equivalent circuit graphs using NetworkX~\cite{SciPyProceedings_11}.  \changes{Our HT benchmarks and test vectors are available to download from \cite{githubGitHubNMSUPEARLHardwareTrojanInsertionandDetectionwithReinforcementLearning}. The HTs are in structural Verilog format, making them easy to use. The input orders of the test vectors are the same as \cite{iscas85}}. Our toolset is developed in Python to 1) quickly adopt available libraries and 2) facilitate future expansions and integration with other tools that researchers may develop. 

Table~\ref{tab:circuits} provides details of the benchmark circuits used in our experiments. The table represents the number of primary inputs ($2^{nd}$column), logic levels ($3^{rd}$column), number of nodes including inputs, outputs, and logic gates ($4^{th}$column), and nets ($5^{th}$column). We have specified $T_{OCR}$ and $T_{HTS}$ such that $5\%$ of all nets in each circuit are considered as \emph{rare nets} ($6^{th}$ and $7^{th}$ columns, respectively). This was done to enable a fair comparison between the circuits. Finally, the circuit functionality is listed in the $8^{th}$ column.
\subsection{Timing Complexity and Scalability}
Table~\ref{tab:Insertion_Detection_complexity} provides timing information on training the HT insertion and detection agents per circuit. The $2^{nd}$ column shows the total timesteps for insertion/detection, and the $3^{rd}$ column shows the total spent time.
We initialize training the inserting agent in $c432$ with $120$K timesteps and an episode length of $450$. We increase both values by 10\% for each succeeding circuit to ensure enough exploration is made in each circuit as their size grows. As for detection, we start with $450$K timesteps and increase it by 10\% for subsequent circuits, and we keep the episode length at 10. The short episode length allows the agent to experience different states, thereby increasing the chances of exploration.
The test vectors are collected after running the agent for $20$K episodes in the testing phase. 

In our experiments, $c6288$ takes the most time in both insertion and detection scenarios (2.5 days), which we argue is reasonable for an attacker and the defense engineer. Note that we have not used optimization techniques to reduce the number of gates and nets in the benchmarks. Such techniques can notably decrease the RL environment size and, subsequently, the training time. That being said, the impact of optimization techniques on detection/insertion quality should be investigated, but it is beyond the scope of this paper.


\begin{table}[!t]
\caption{Mean HT detection/insertion training time of the RL algorithm  for different ISCAS-85 benchmarks}
\label{tab:Insertion_Detection_complexity}
\begin{tabular}{|c|c|c|}
\hline
Benchmark& \centered{Insertion/Detection \\ ~~~~~Timesteps} & \centered{Insertion/Detection \\ ~~~Training Time} \\ \hline
 \centered{c432}& 120K / 450K & 1 hr 40 m / 1 hr 7 m\\\hline
 \centered{c880}& 132K / 495K & 2 hr 36 m / 2 hr 7 m \\\hline
 \centered{c1355}& 145K / 550K & 3 hr 10 m / 2 hr 27 m  \\\hline
 \centered{c1908}& 160K / 605K & 5 hr 25 m / 2 hr 40 m  \\\hline
 \centered{c2670}& 175K / 665K & 8 hr 1 m / 7 hr 23 m  \\\hline
 \centered{c3540}& 192K / 731K & 12 hr 1 m / 5 hr 24  \\\hline
 \centered{c5315}& 211K / 800K & 23 hr 16 m / 15 hr 36 m  \\\hline
 \centered{c6288}& 232K / 880K & 57 hr 18 m / 59 hr 16 m  \\\hline
 \centered{c7552} & 255K / 970K & 26 hr 15 m / 44 hr 15 m \\\hline
\end{tabular}
\end{table}

\begin{figure*}[!t]
     \centering
     \begin{subfigure}[b]{0.75\textwidth}
         \centering
         \includegraphics[width=\textwidth]{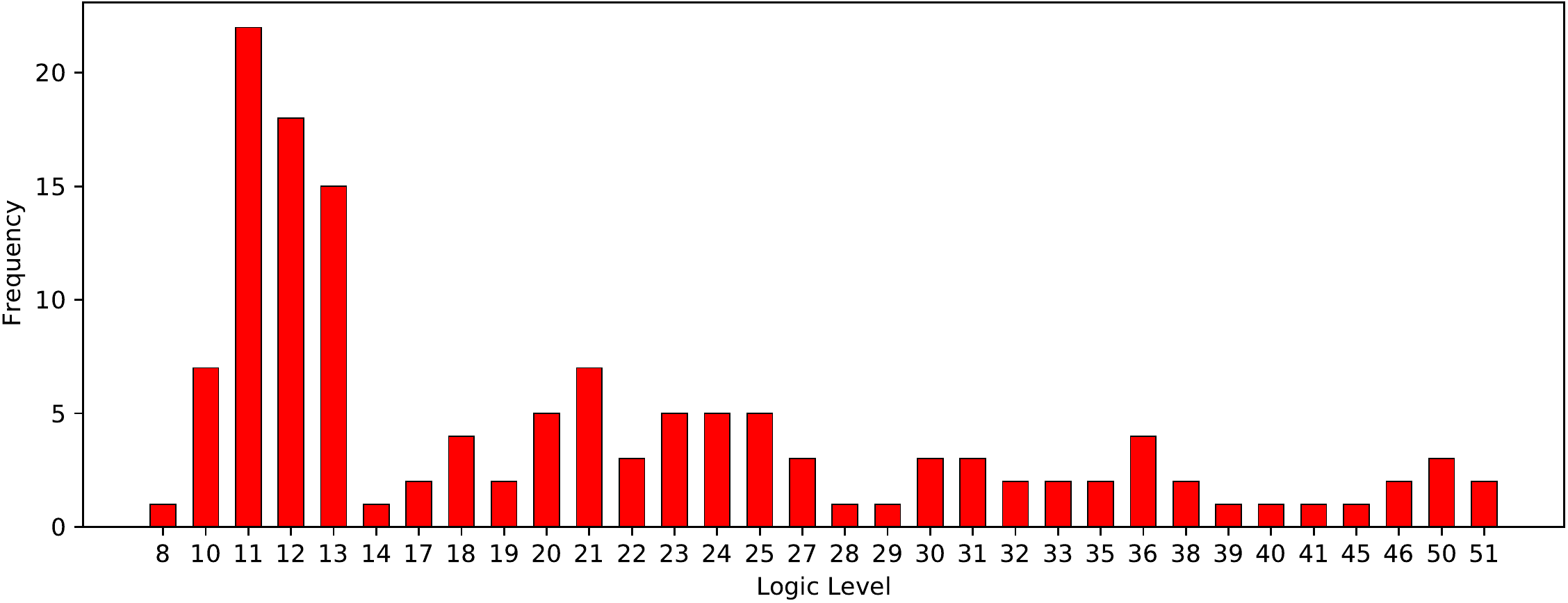}
         \caption{c3540}
         \label{c3540_rare_nets}
     \end{subfigure}
      \begin{subfigure}[b]{0.75\textwidth}
         \centering
         \includegraphics[width=\textwidth]{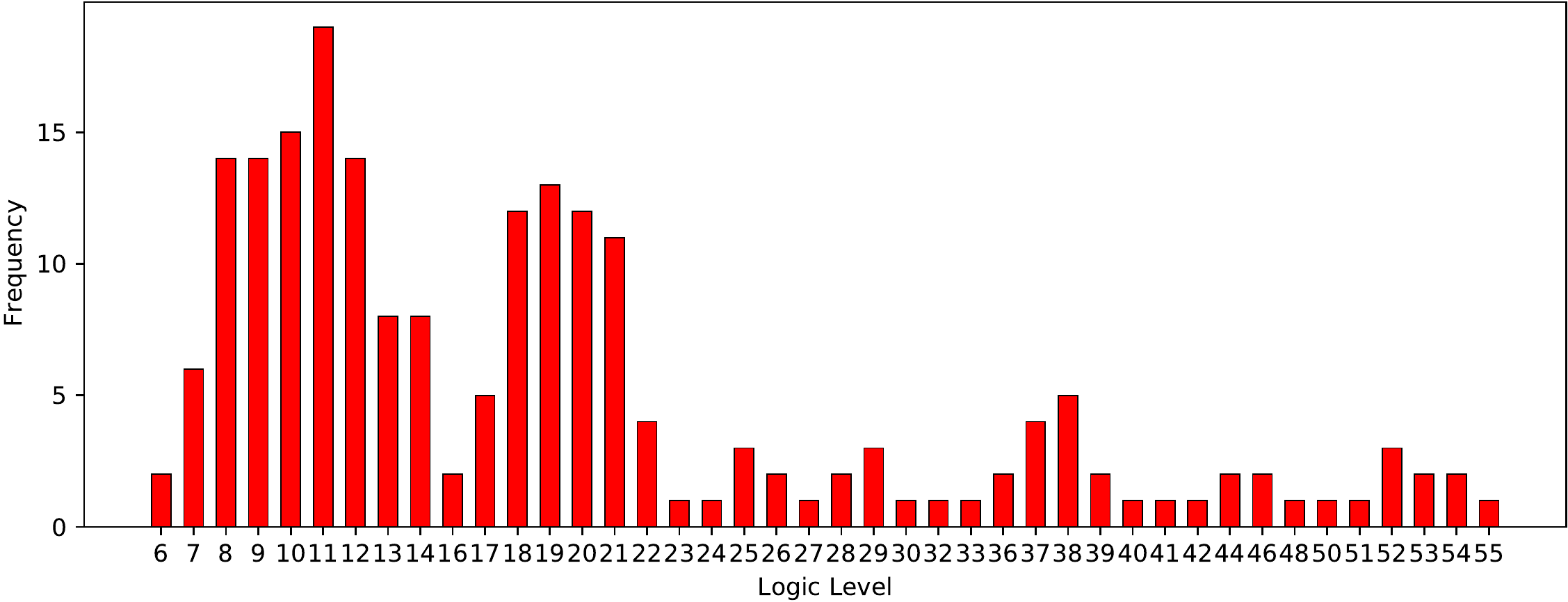}
         \caption{c5315}
         \label{c5315_rare_nets}
     \end{subfigure}
        \caption{Distribution of rare nets in $c3540$ and $c5315$}
        \vspace{-6mm}
        \label{fig:rare_net_distribution}
\end{figure*}

\begin{table*}[!b]
\caption{Number of inserted HTs under $P_{rand}$ and $P_{high}$ scenarios for ISCAS-85 benchmark circuits}
\label{tab:Insertion}
\begin{center}
 \scalebox{0.99}{
\begin{tabular}{|c||c|c||c|c||c|c||c|c|}
\hline
\textbf{Benchmark}&  $P_{rand}- Total$ & $P_{high}- Total$ & $P_{rand}-3 $ & $P_{high}-3 $& $P_{rand}-4 $ & $P_{high}-4 $& $P_{rand}-5 $ & $P_{high}-5$  \\ \hline
   \textbf{c432}&1866&2788&1688&2331&160&453&18&4\\\hline
   \textbf{c880}&1954&2116&1595&1736&327&373&32&7\\\hline
   \textbf{c1355}&921&1400&815&1116&86&268&20&16\\\hline
   \textbf{c1908}&1247&1576&1121&1240&126&321&\textbf{0}&15\\\hline
   \textbf{c2670} &206&434&188&406&18&28&\textbf{0}&\textbf{0}\\\hline
   \textbf{c3540}&410&767&367&703&41&64&2&\textbf{0}\\\hline
   \textbf{c5315}&434&797&406&719&28&77&\textbf{0}&1\\\hline
   \textbf{c6288}&531&475&459&426&67&46&5&3\\\hline
   \textbf{c7552}&769&683&704&615&64&67&1&1\\\hline
\end{tabular}
}
\end{center}
\end{table*}


\begin{figure*}[!t]
     \centering
     \begin{subfigure}[b]{0.75\textwidth}
         \centering
         \includegraphics[width=\textwidth]{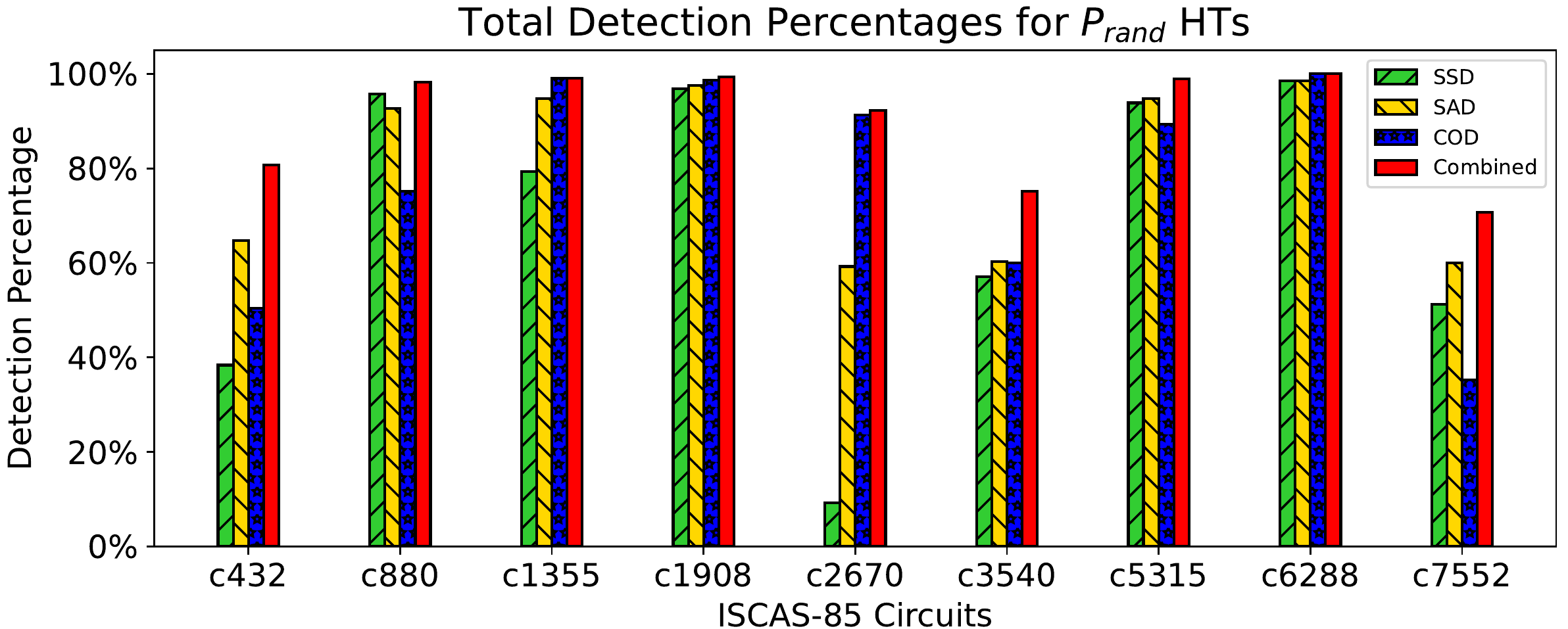}
         \caption{$P_{rand}$ HTs}
     \end{subfigure}
      \begin{subfigure}[b]{0.75\textwidth}
         \centering
         \includegraphics[width=\textwidth]{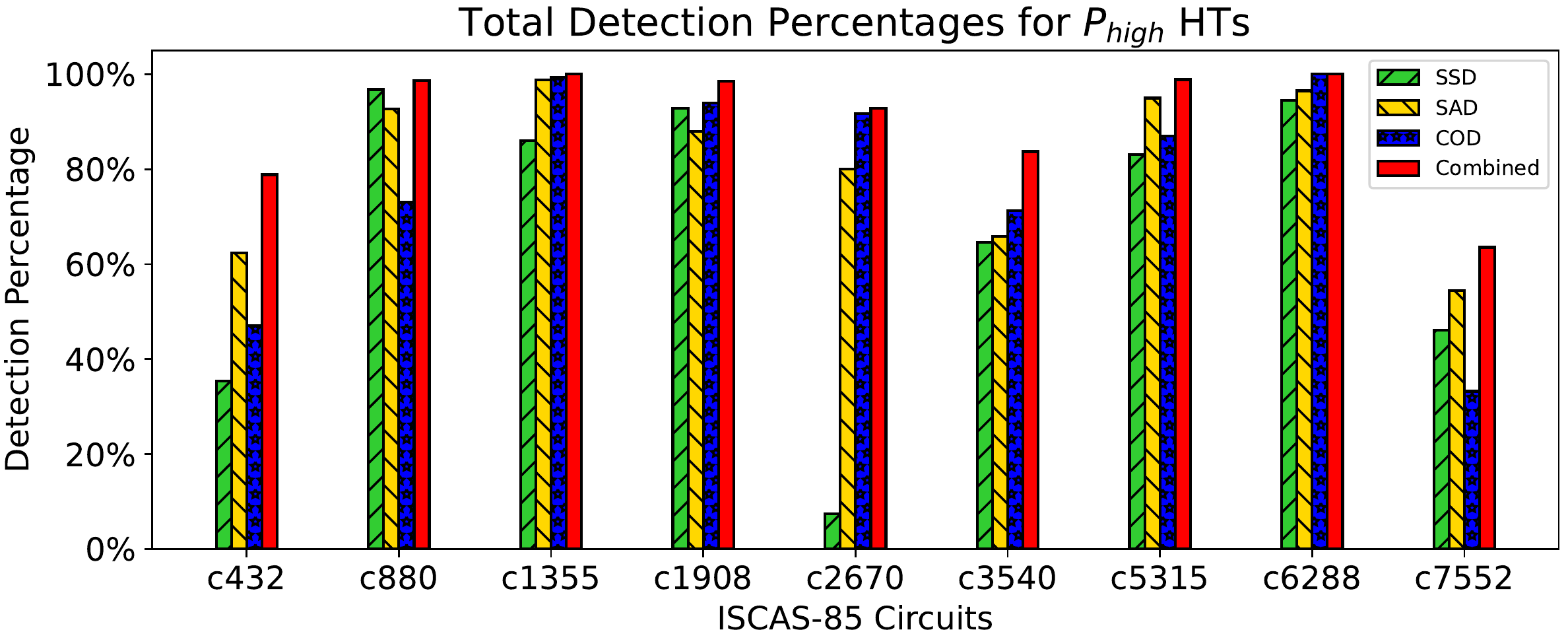}
         \caption{$P_{high}$ HTs}
     \end{subfigure}
     \caption{Detection accuracy of SSD, SAD, COD, and \textit{Combined} scenarios under $P_{high}$ and $P_{high}$ insertion scenarios in ISCAS-85 benchmark circuits }
     \vspace{-6mm}
     \label{fig:detection_graphs}
\end{figure*}

 \begin{figure*}[!t]
 \centering
      \begin{subfigure}[b]{0.75\textwidth}
     \centering
     \includegraphics[width=\textwidth]{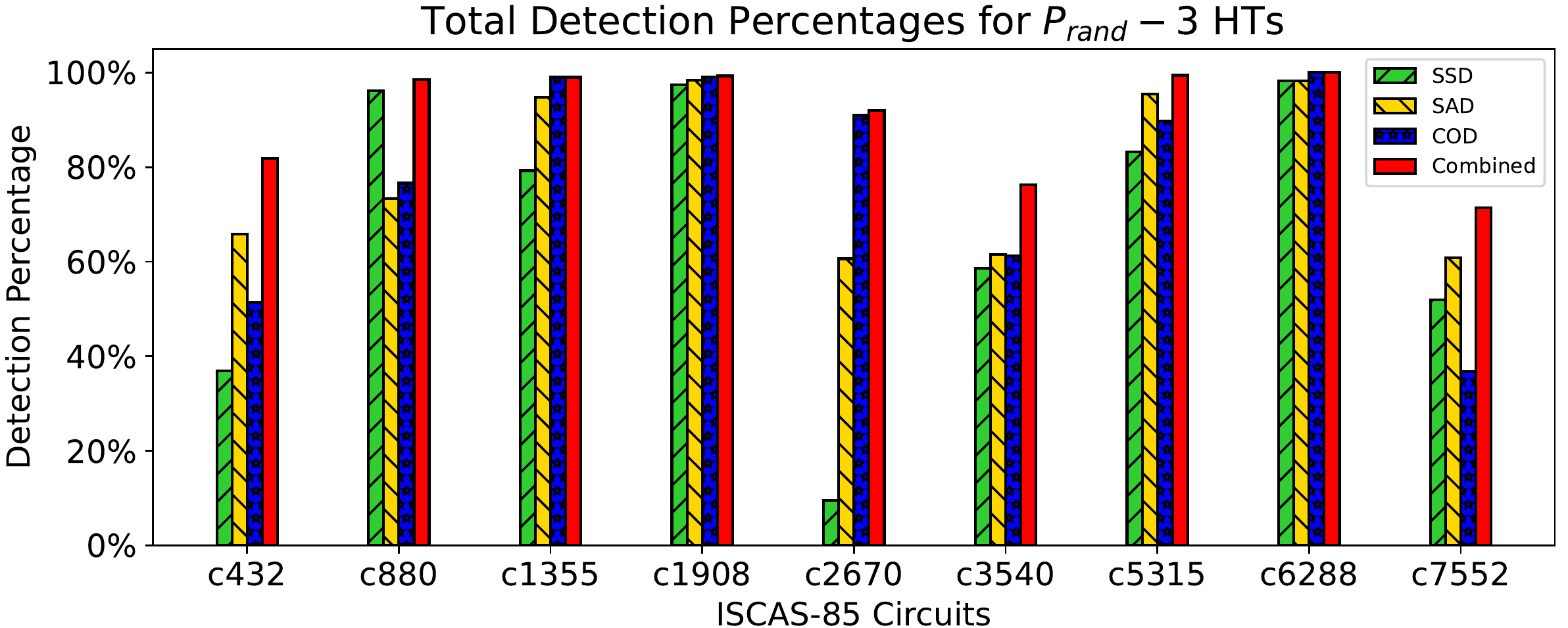}
     \caption{$P_{rand}-3~HTs$}
     \end{subfigure}
      \begin{subfigure}[b]{00.75\textwidth}
         \centering
         \includegraphics[width=\textwidth]{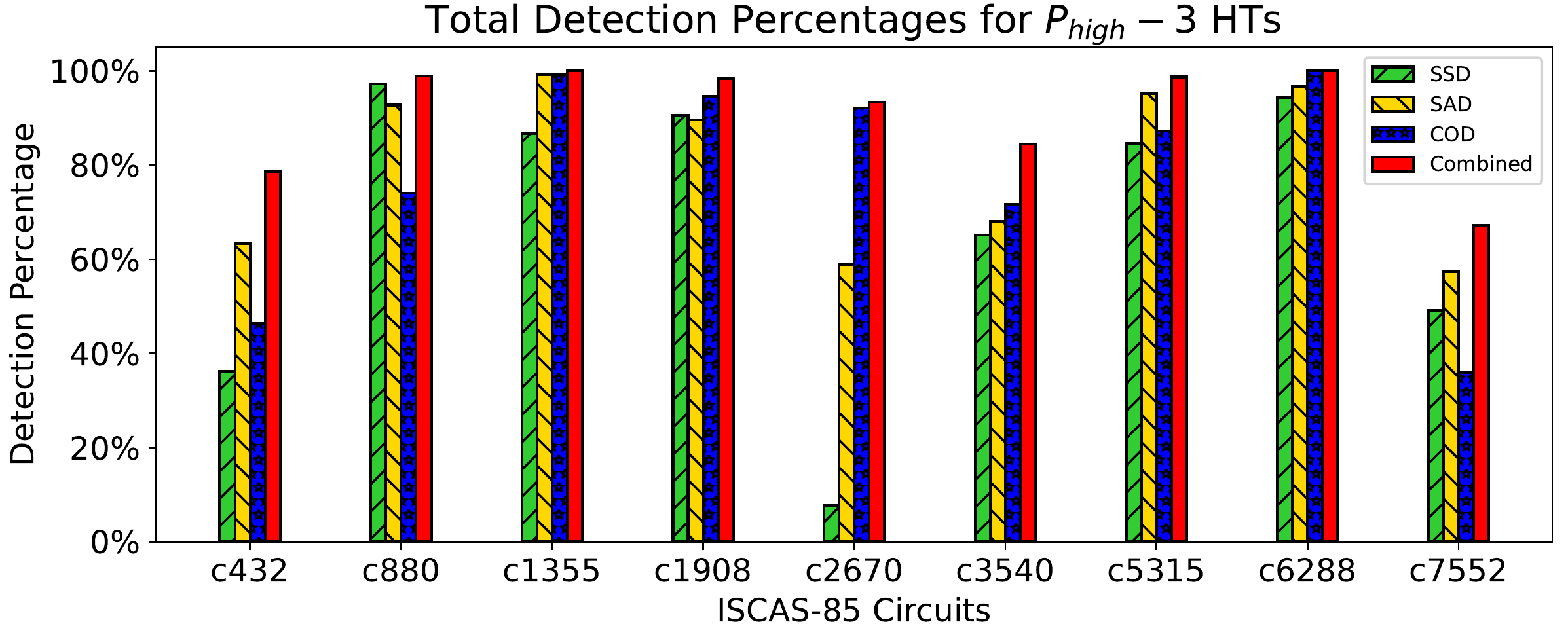}
         \caption{$P_{high}-3~HTs$}
     \end{subfigure}
     \caption{Detection accuracy of SSD, SAD, COD, and \textit{Combined} scenarios under $P_{high}$ and $P_{high}$ insertion scenarios for 3-input HTs}
     \vspace{-6mm}
     \label{fig:detection_graphs3}
\end{figure*}

\begin{figure*}[!t]
 \centering
     \begin{subfigure}[b]{0.75\textwidth}
     \centering
     \includegraphics[width=\textwidth]{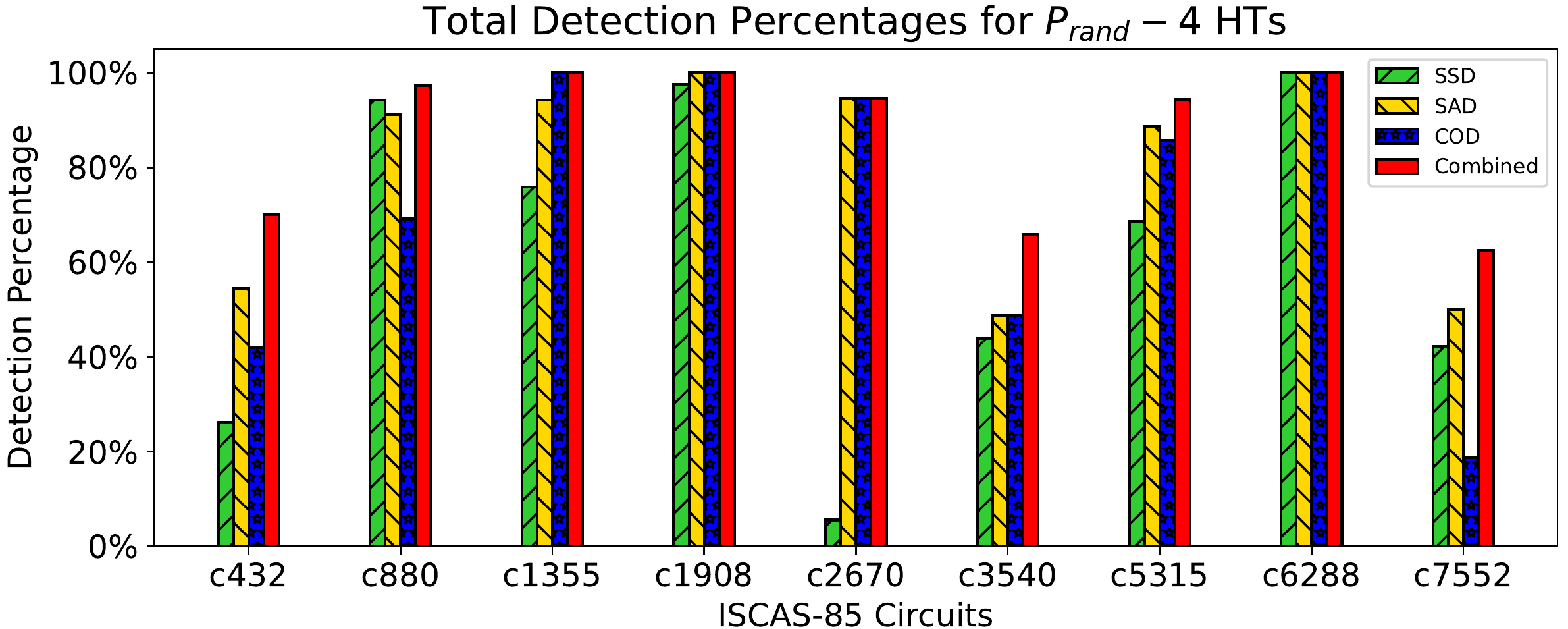}
     \caption{$P_{rand}-4~HTs$}
     \end{subfigure}
      \begin{subfigure}[b]{0.75\textwidth}
         \centering
         \includegraphics[width=\textwidth]{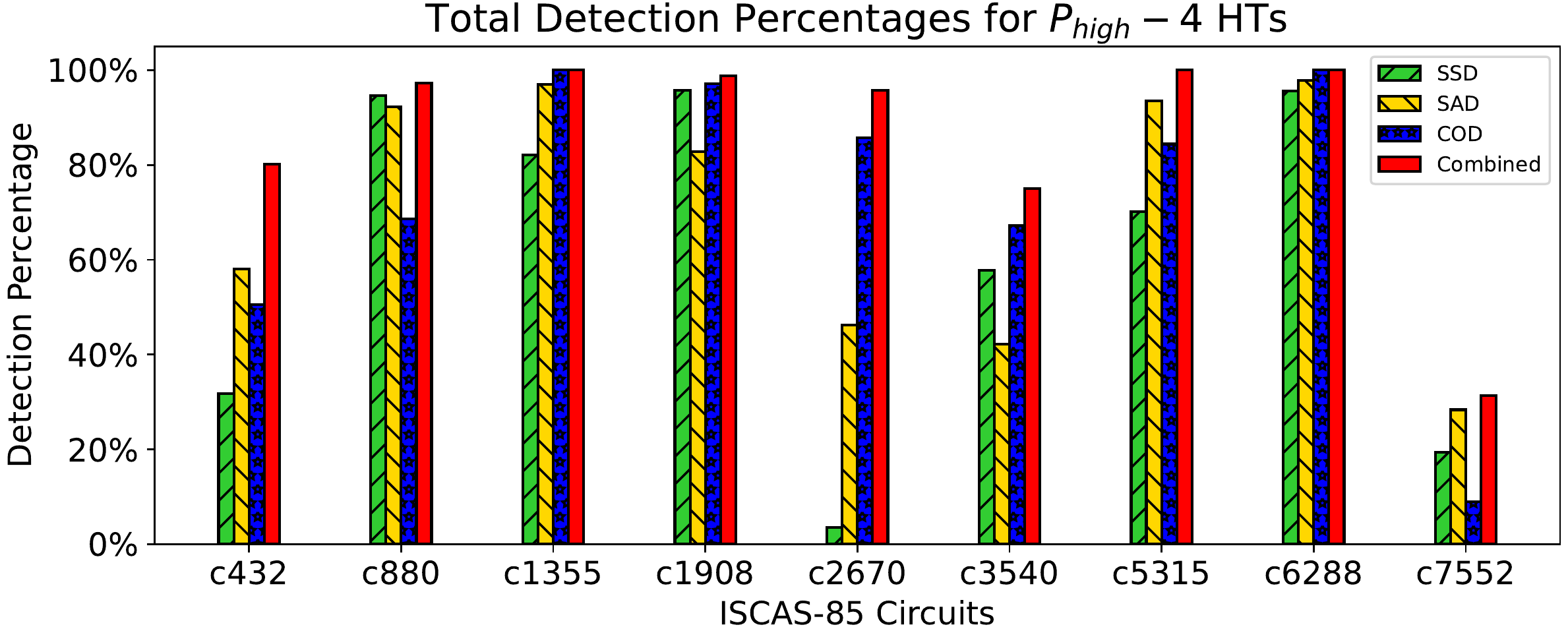}
         \caption{$P_{high}-4~HTs$}
     \end{subfigure}
     \caption{Detection accuracy of SSD, SAD, COD, and \textit{Combined} scenarios under $P_{high}$ and $P_{high}$ insertion scenarios for 4-input HTs}
     \vspace{-6mm}
     \label{fig:detection_graphs4}
\end{figure*}
\begin{figure*}[!t]
 \centering
    \begin{subfigure}[b]{0.75\textwidth}
     \centering
     \includegraphics[width=\textwidth]{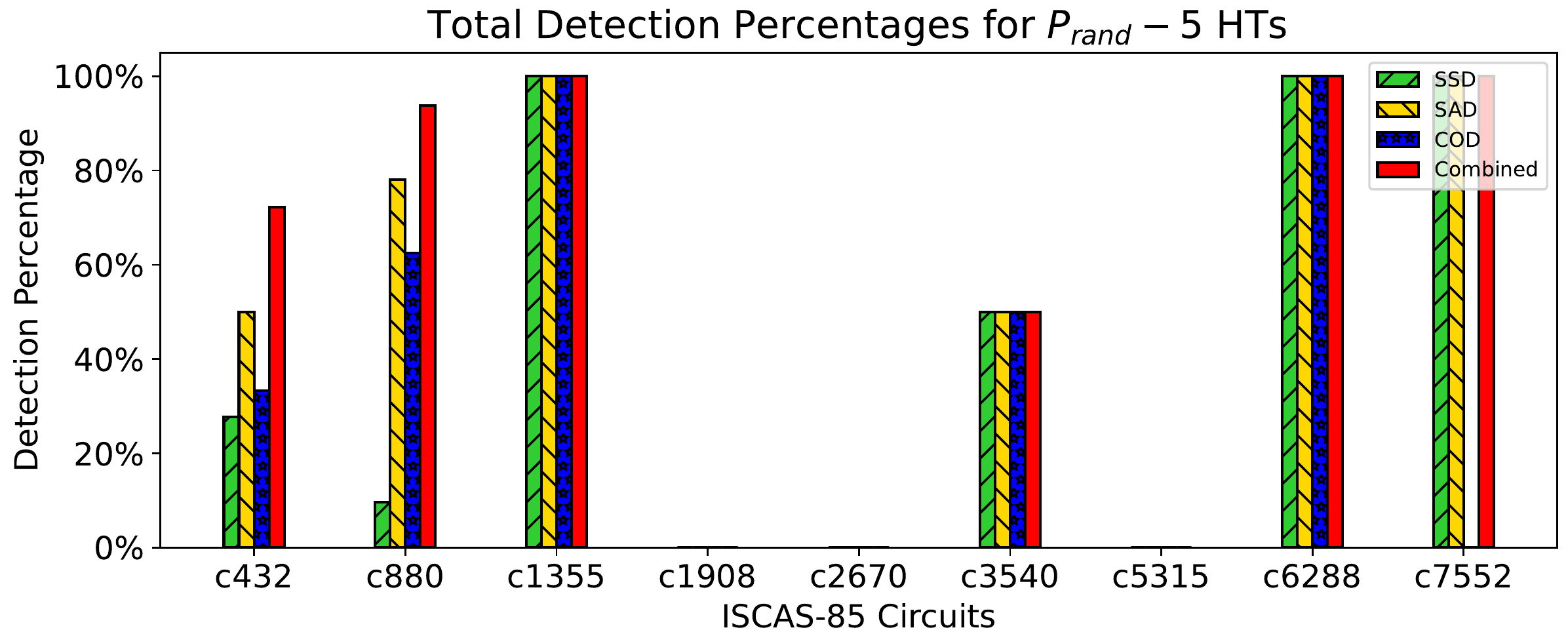}
     \caption{$P_{rand}-5~HTs$}
     \end{subfigure}
      \begin{subfigure}[b]{0.75\textwidth}
         \centering
         \includegraphics[width=\textwidth]{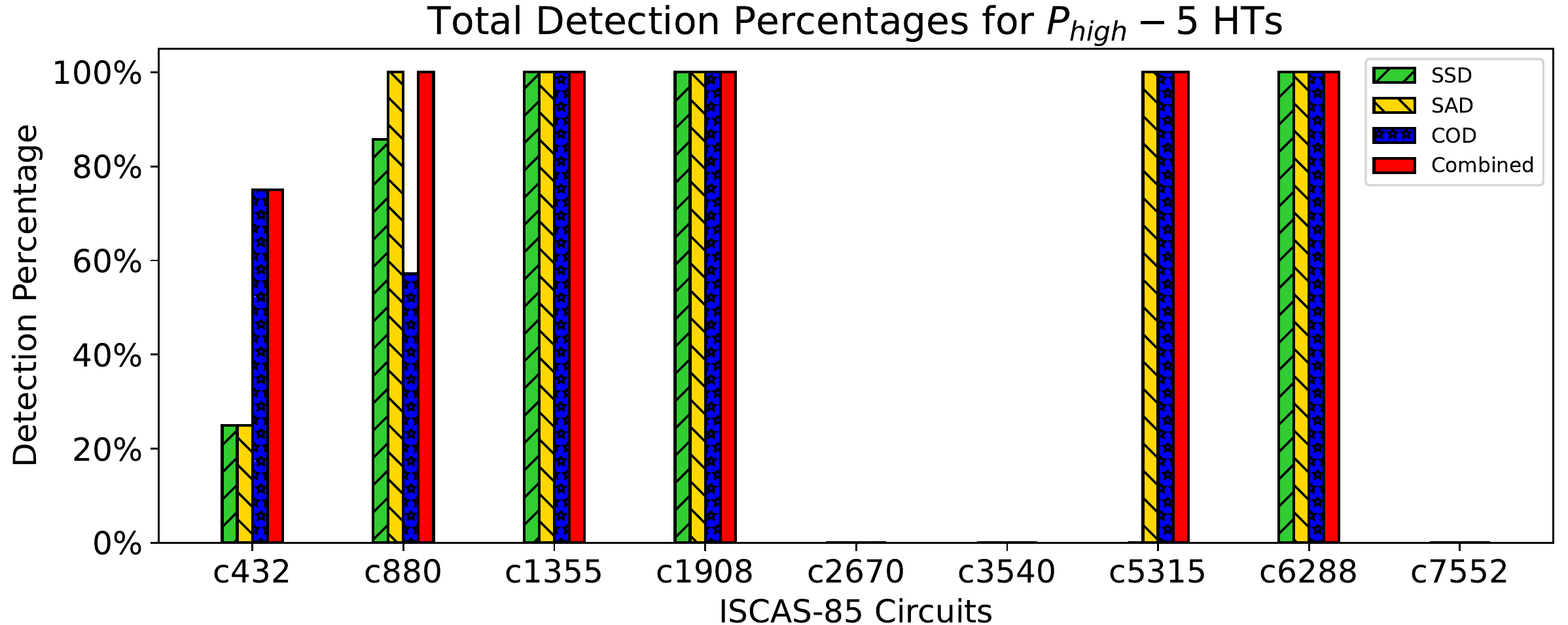}
         \caption{$P_{high}-5~HTs$}
     \end{subfigure}
\caption{Detection accuracy of SSD, SAD, COD, and \textit{Combined} scenarios under $P_{high}$ and $P_{high}$ insertion scenarios for 5-input HTs}
    \label{fig:detection_graphs5}
        \vspace{-5mm}
\end{figure*}

\subsection{Insertion, Detection, and Confidence Value Figures}

Figure~\ref{fig:rare_net_distribution} illustrates the logical depth distribution of rare nets in $c3540$ and $c5315$ circuits. 
Although rare nets are primarily found in the lower logic levels, there are still a significant number of rare nets in the higher levels, which could contribute to the creation of stealthier hardware Trojans. As explained in section~\ref{subsec:RL_HT_insertion}, the level of the HT trigger nets is limited by the payload's level. Suppose a payload is not selected from the higher-level nets. In that case, the agent has less opportunity to explore higher-level trigger nets, which might harm the insertion exploration of new HTs. To enable more exploration, we define the following two payload selection scenarios: 1) \textit{$P_{rand}$} in which the agent selects payloads randomly, and 2) \textit{$P_{high}$} where payload net is selected such that at least 80\% of rare nets are within the agent's sight. 
 
Table~\ref{tab:Insertion} provides information about the number of inserted HTs using $P_{rand}$ and $P_{high}$ scenarios for each benchmark circuit. The \nth{2} and \nth{3} columns show the total number of HTs successfully inserted by the agent. The numbers followed by each insertion scenario in the remaining columns show the number of rare nets among the five input triggers. For instance, in $c432$, $1866$ HTs were inserted under $P_{rand}$ where $1688$ of those had $3$ rare nets, $160$ of those had $4$ rare nets, and only $18$ of those had $5$ rare nets. As can be observed, in most cases, the number of inserted HTs under $P_{high}$ is higher than $P_{rand}$ except for $c6288$ and $c7552$. Also, fewer HTs are inserted as the number of rare triggers increases. In other words, it becomes more difficult for the RL agent to find HTs with higher rare nets. There are some cases under $P_{rand}-5 $ and $P_{high}-5$ that the agent could not insert any HTs. These rows in the table are shown as $0$, \eg, in $c2670$.

\begin{figure}[!t]
  \centering
  \includegraphics[width=.49\textwidth]{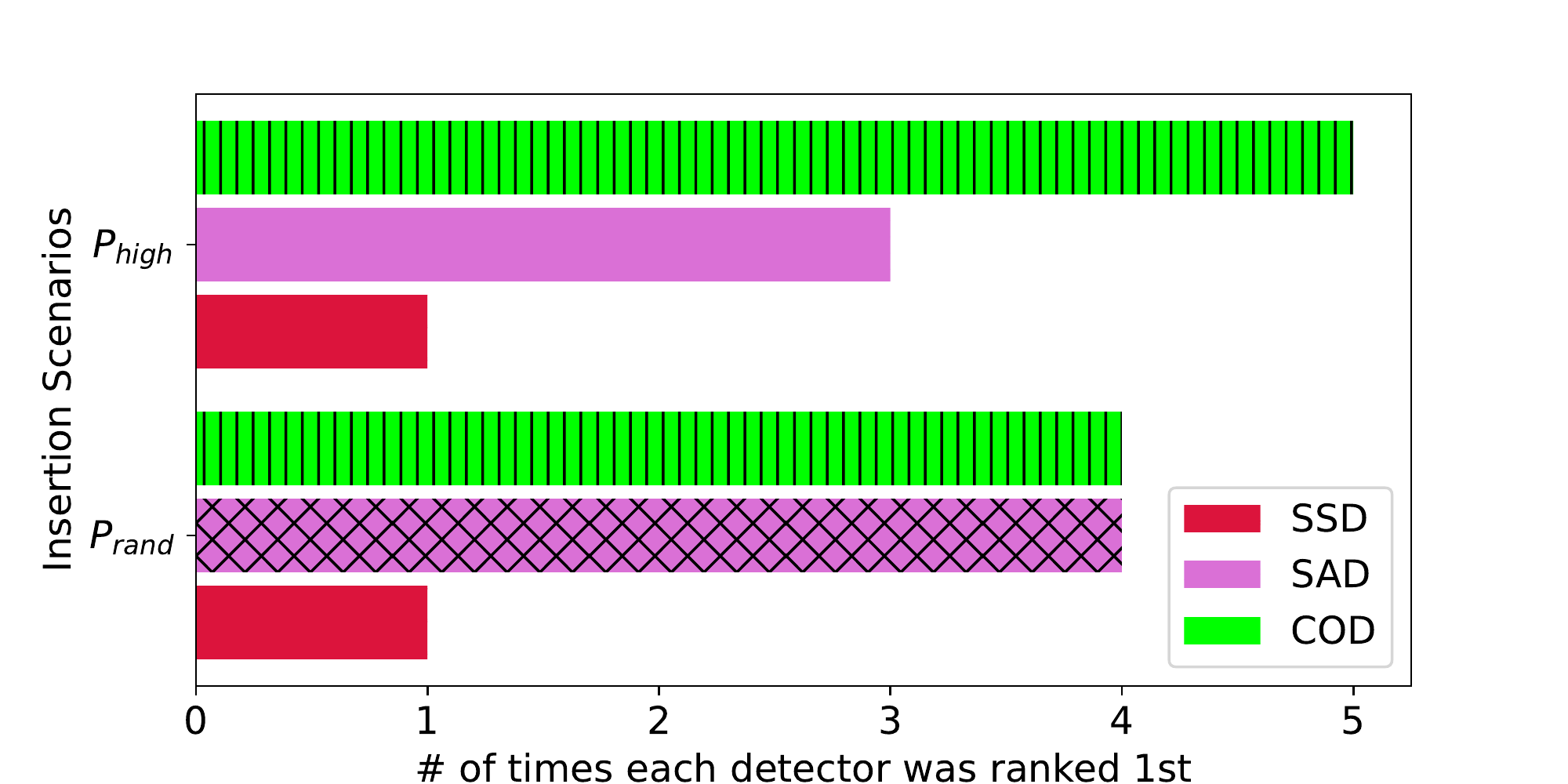}
  \caption{Comparing the number of times each of $SSD$, $SAD$, and $COD$ are ranked as the best detector in our two insertion scenarios}
  \vspace{-5mm}
  \label{fig:ranking}
\end{figure}

\changes{Figure~\ref{fig:detection_graphs} displays the HT detection accuracy percentages for the studied circuits under $P_{rand}$ and $P_{high}$ insertion scenarios. Figures ~\ref{fig:detection_graphs3},~\ref{fig:detection_graphs4}, and~\ref{fig:detection_graphs5} provide details about the detection accuracy of each HT group, separately.} Besides $SSD$, $SAD$, and $COD$, there is an extra detection scenario called \textit{Combined} where all the test vectors produced by $SSD$, $SAD$, and $COD$ are consolidated and applied to the circuits for HT detection. No detection rates are reported in cases where no HTs were inserted. It can be observed from both Table~\ref{tab:Insertion} and Figures~\ref{fig:detection_graphs},~\ref{fig:detection_graphs3},~\ref{fig:detection_graphs4}, and~\ref{fig:detection_graphs5} that despite more inserted HTs in the $P_{high}$ scenario, they do not evade detection any better than the random payload selection scenario and the detection rates are almost the same. Nevertheless, the extra inserted HTs under $P_{high}$ can be used to train better ML HT detectors. Figures ~\ref{fig:detection_graphs},~\ref{fig:detection_graphs3},~\ref{fig:detection_graphs4}, and~\ref{fig:detection_graphs5} also suggest that SSD, SAD, and COD are vital to providing better HT detection coverage. Figure~\ref{fig:ranking} displays the number of times each detector was ranked first in nine benchmark circuits under our two insertion strategies. While $COD$ ties with $SAD$ under $P_{rand}$, it becomes the best detector under $P_{high}$. $SSD$ only outperforms in 1 benchmark circuit in both scenarios. The figure suggests that solely developing HT detectors based on signal activity might not achieve the expected outcomes. Nevertheless, $SAD$ still plays an essential role in overall HT detection accuracy. The impact of the \textit{Combined} scenario is vital as it improves the overall detection accuracy in most cases. For instance, in $c3540$, none of the detectors can perform better than $60\%$ in the $P_{rand}$ scenario while the \textit{Combined} detection accuracy is nearly $75\%$. It also can be seen that adding more rare nets to the HT trigger does not necessarily lead to stealthier HTs. For example, in $c880$, $c1355$, and $c1908$, there are HTs with five trigger nets that were 100\% detected, while the detection accuracy was less for HTs with fewer rare triggers in the same circuits.

Another important observation is the different magnitude of detection accuracy among the benchmark circuits. While we achieve $100\%$ accuracy in $c6288$, it is about $25\%$-$30\%$ lower in $c3540$ and $c6288$. Table ~\ref{tab:circuits} shows that $c6288$ is a multiplier circuit. It contains $240$ full and half adders arranged in a $15\times16$ matrix~\cite{iscas85}. $c3540$, on the other hand, has $14$ control inputs for multiplexing and masking data. $c7552$ also contains multiple control signals and bit masking operations. We hypothesize that the detection accuracy is higher in $c6288$ due to having fewer control signals that disable circuit components and signals. Accordingly, they get more frequently activated in $c6288$ than $c3540$ and $c7552$. \textbf{In other words, these results imply that inserting HTs in control paths can lead to stealthier HTs than data paths in circuits.} Another interesting finding pertains to the detection rate in $c432$. After administering $100$K random test patterns, we discovered that the rarest net in the circuit was triggered  $7\%$ of the times, starkly contrasting to other circuits where many nets exhibit less than $1\%$ switching activity. It implies that random test patterns probably more easily activate the inserted HTs in $c432$. We generated $20$K random test patterns to prove this hypothesis and passed them to the circuit. These test patterns detected $99\%$ of HTs, indicating that attackers should carefully evaluate the activity profile of the nets before compromising circuits.

\begin{figure*}[!t]
     \centering
     \begin{subfigure}[b]{0.6\textwidth}
         \centering
         \includegraphics[width=\textwidth]{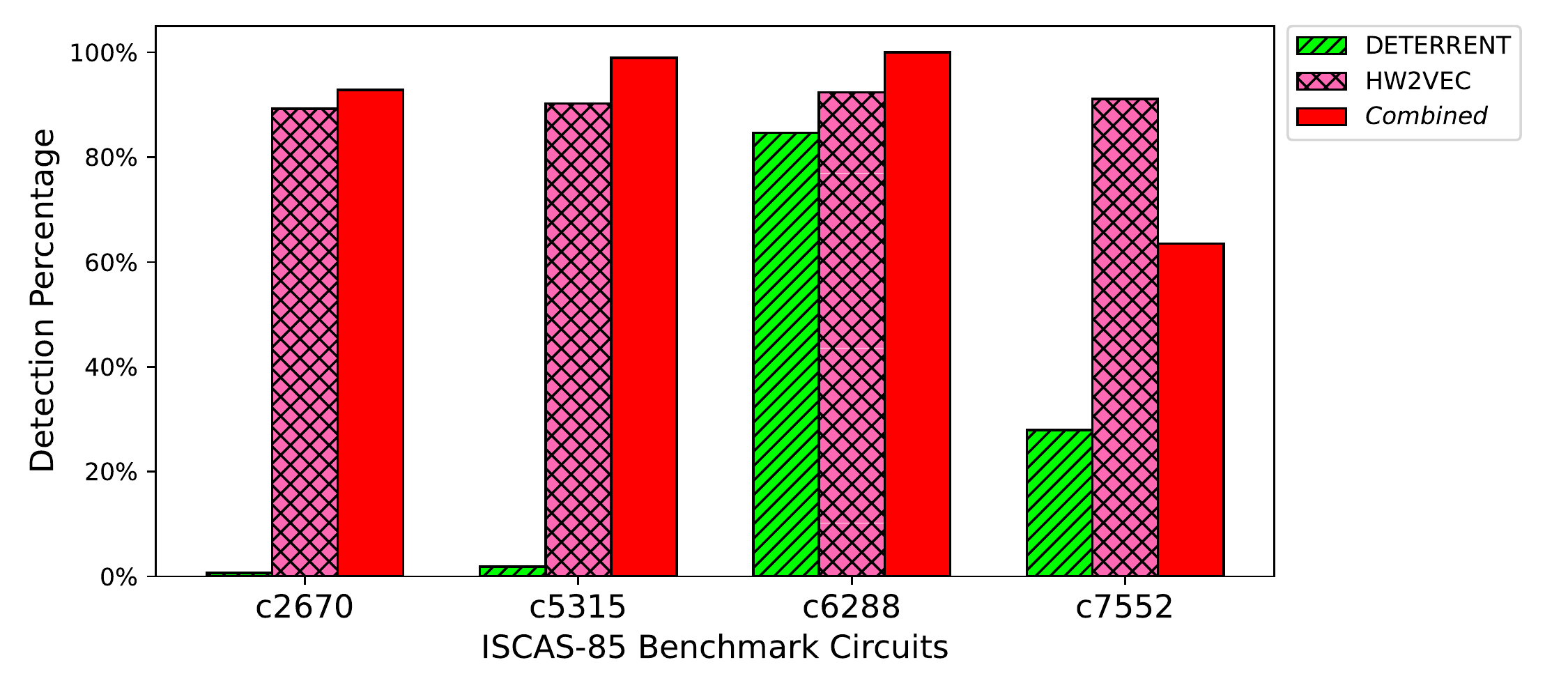}   
         \vspace{3mm}
         \caption{$P_{rand}~HTs$ }
     \end{subfigure}
     \hfill
     \centering
      \begin{subfigure}[b]{0.6\textwidth}
         \centering
         \includegraphics[width=\textwidth]{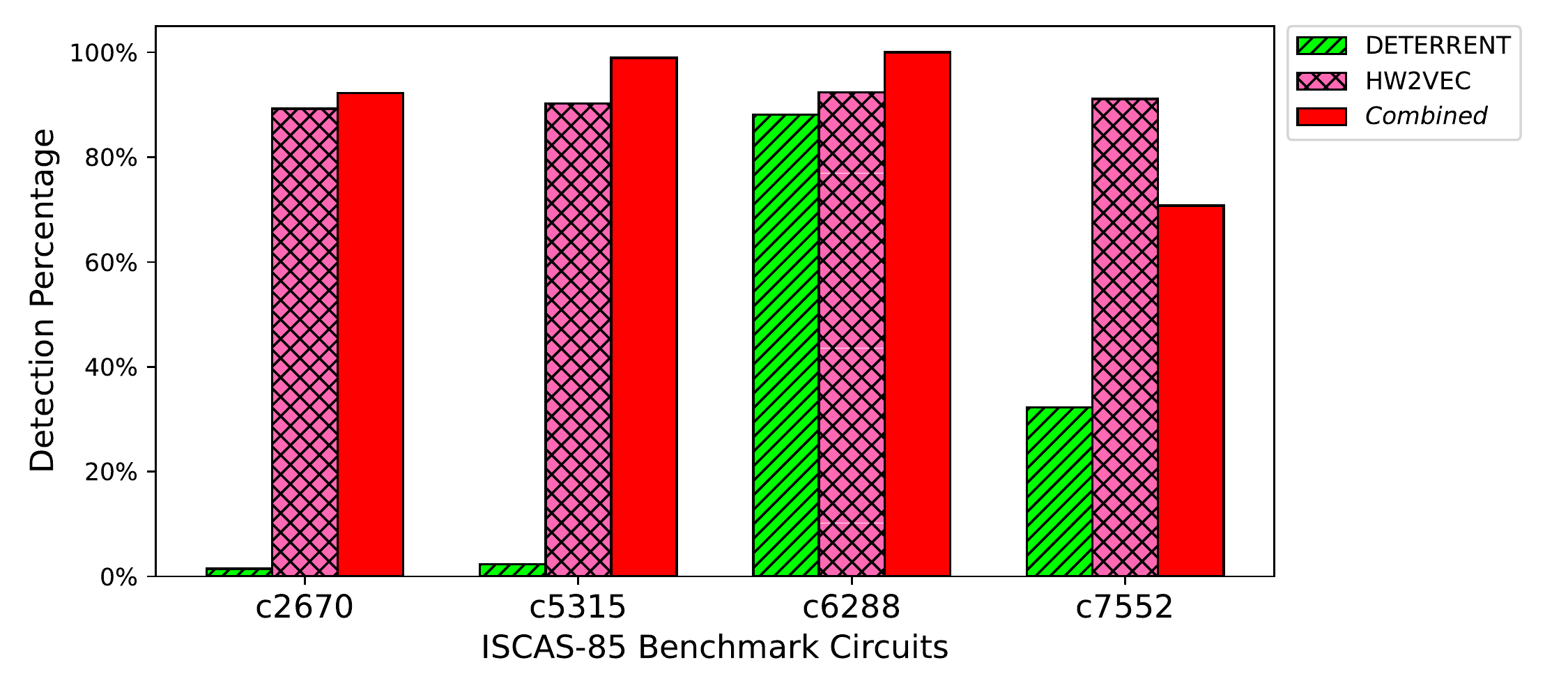}
         \caption{$P_{high}~HTs$}
     \end{subfigure}

\caption{Comparison of HW2VEC~\cite{yu2021hw2vec}, \textit{Combined},  and DETERRENT~\cite{gohil2022deterrent} detection rates under (a) $P_{rand}$ and (b) $P_{high}$ insertion scenarios}
\vspace{-6mm}
\label{fig:DETERRENT_HW2VEC}
\end{figure*}

To further evaluate the efficacy of our HT detectors, we compare the \textit{Combined} detector with DETERRENT~\cite{gohil2022deterrent} and HW2VEC~\cite{yu2021hw2vec}, two state-of-the-art HT detectors. We use the test vectors generated by DETERRENT~\cite{gohil2022deterrent} and collect detection figures for $4$ reported ISCAS-85 benchmark circuits, namely $c2670$, $c5315$, $c6288$, and $c7552$\footnote{We reached out to the authors of TARMAC and TGRL techniques, but we did not receive the test patterns at the time of submission.}. We also replicate the steps in HW2VEC~\cite{yu2021hw2vec} by gathering the $TJ\_RTL$ dataset, which contains $26$ HT-infected (labeled as $'1'$) and $11$ HT-Free circuits (labeled as $'0'$). We train an MLP (multi-layer perceptron) binary classifier to detect the HTs. For the test dataset, we collect the graph embeddings of the HTs generated by the inserting RL agent. Additionally, we add an HT-free version of the original ISCAS-85 circuits and another one synthesized with the academic \textit{NanGateOpenCell45nm} library to the test batch to record the number of $TN$s and $FP$s. As shown in Table~\ref{tab_detection}, DETERRENT solely considers signal activity while HW2VEC captures structural information of circuits.

 Figure~\ref{fig:DETERRENT_HW2VEC} shows the detection accuracy of each HT detector for each benchmark circuit. The detection accuracy is reported for the total inserted HTs in Table~\ref{tab:Insertion} for both $P_{rand}$ and $P_{high}$ insertion scenarios. The figure shows that the \textit{Combined} detector outperforms DETERRENT and HW2VEC in $3$ of our benchmark circuits. The average detection rate among the $4$ benchmarks is $87\%$ percent. While the detection gap between \textit{Combined} and DETERRENT is significant in $c2670$ and $c5315$, it is less evident in $c6288$ and $c7552$. HW2VEC, on the other hand, demonstrates minimal detection variance in all $4$ circuits and outperforms \textit{Combined} in $c7552$. Furthermore, HW2VEC illustrates robust performance with HT-Free circuits, correctly classifying them as $TN$s and a $FP$ rate of $0$. 

In another experiment, we train our MLP with $TJ\_RTL$ $+$ $EPFL$~\cite{amaru2015epfl} benchmark suites to obtain a more balanced dataset ($26$ instances labeled as $'1'$ and $30$ instances labeled as $'0'$). While the $FP$ remains $0$, similar to the previous experiment, the HT detection accuracy drops to $48\%$. This sheds light on the shortcomings of the current benchmarks used for training ML HT detectors, and it raises the necessity of having a more diverse and larger dataset to attain more dependable results. Overall, these two experiments demonstrate the potential of the RL inserting agent and the advantages of a multi-criteria detector compared to a single-criterion (DETERRENT) HT detector.

Table~\ref{tab:detection_contribution} shows the individual detection contribution of $SSD$, $SAD$, and $COD$ towards overall HT detection for each benchmark circuit. The $2^{nd}$, $4^{th}$ and $6^{th}$ columns display the number of HTs exclusively detected by each detector followed by their contribution in the overall HT detection in the $3^{rd}$, $5^{th}$ and $7^{th}$ columns for $SSD$, $SAD$, and $COD$, respectively. As can be inferred, $COD$ has the highest individual contribution, followed by $SAD$ and $SSD$. This table is evidence of the importance of the multi-criteria HT detector for higher accuracy.

To compute the confidence value of each detector, the overall detection accuracy of each detector is calculated in all nine circuits under both insertion scenarios. Then, each averaged value is plugged into Equation~\ref{equ:metric}. Assuming $\alpha = 10$, the confidence values for each $SSD$, $SAD$, $COD$, and \textit{combined} scenarios are 2.43, 3.36, 3.09, and 5.13, respectively. Thus, the security engineer can put more confidence in the \textit{Combined} detector since it has the highest confidence values. DETERRENT's and HW2VEC's confidence values are 1.24 and 4.34, respectively.

\begin{table}[!t]
    \caption{Individual contribution of $SSD$, $SAD$, and $COD$ in detection of unique HTs}

    \label{tab:detection_contribution}
    \scalebox{0.9}{
        \begin{tabular}{|c|c|c|c|c|c|c|}
            \hline
            \textbf{Circuit} & $SSD$ \#&$SSD$ \%& $SAD$ \#  & $SAD$      \% & $COD$ \#& $COD$ \% \\ \hline
               \textbf{c432}&2&0.1\%&275&14.74\%&297&15.86\%\\\hline
               \textbf{c880}&49&2.52\%&16&0.81\%&16&0.81\%\\\hline
               \textbf{c1355}&0&0\%&0&0\%&40&4.34\%\\\hline
               \textbf{c1908}&1&0.08\%&1&0.08\%&13&1.04\%\\\hline
               \textbf{c2670}&0&0\%&1&0.48\%&66&32.03\%\\\hline
               \textbf{c3540}&7&1.70\%&29&7.07\%&18&4.39\%\\\hline
               \textbf{c5315}&1&0.24\%&8&1.93\%&9&2.17\%\\\hline
               \textbf{c6288}&0&0\%&0&0\%&8&1.51\%\\\hline
               \textbf{c7552}&16&2.08\%&29&3.77\%&15&1.95\%\\\hline
        \end{tabular}
    }

\end{table}

\subsection{Average Episode Length and Reward }
\begin{figure*}[!h]
     \centering
     \begin{subfigure}[b]{0.47\textwidth}
         \centering
         \includegraphics[width=\textwidth]{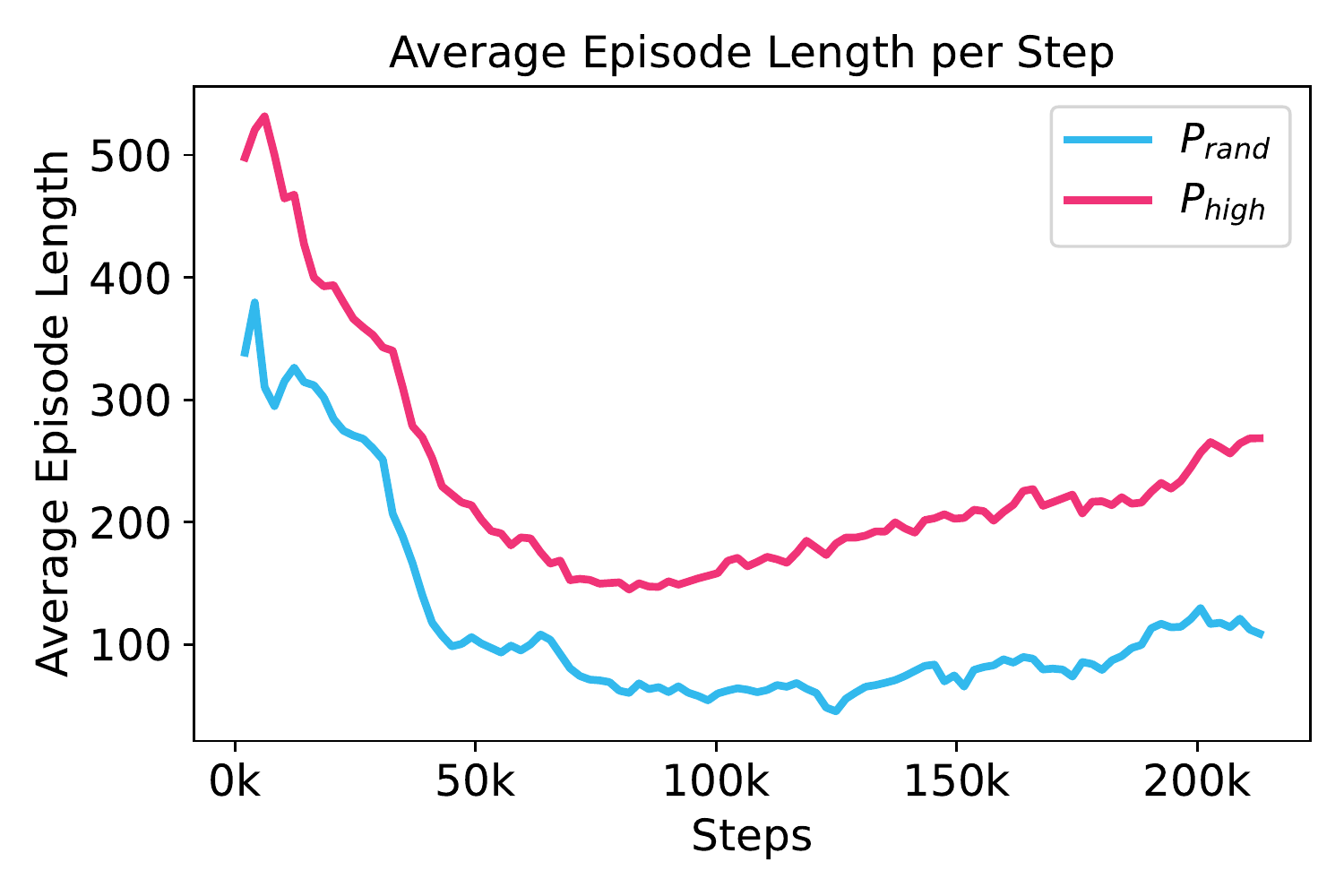}
         
         \caption{Average Episode Length per Step in HT insertion for $c5315$}
         \label{fig:reward_B}
     \end{subfigure}
      \begin{subfigure}[b]{0.47\textwidth}
         \centering
         \includegraphics[width=\textwidth]{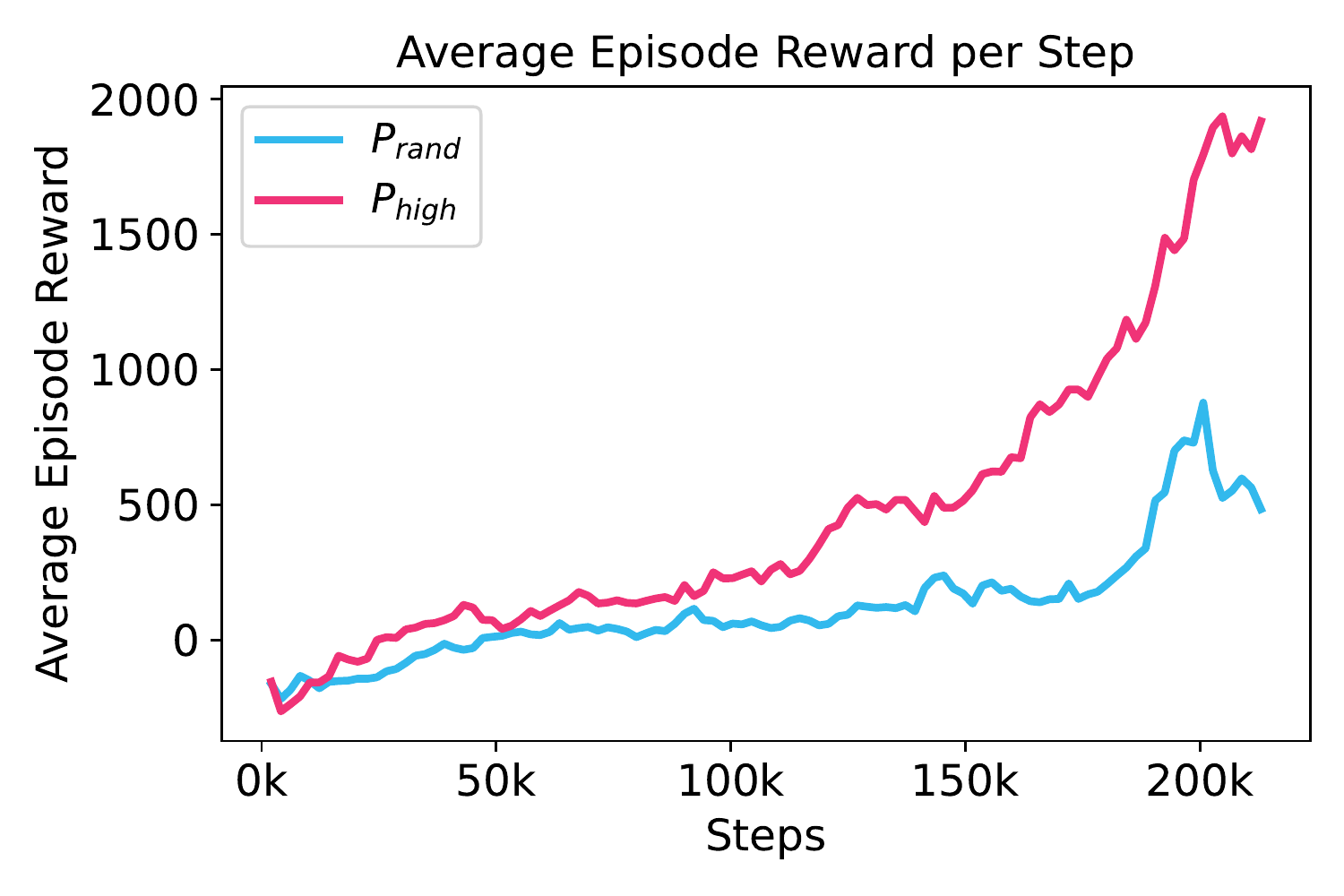}
         \caption{Average Episode Reward per Step in HT insertion for $c5315$}
     \end{subfigure}
    \begin{subfigure}[b]{0.55\textwidth}
         \centering
         \includegraphics[width=\textwidth]{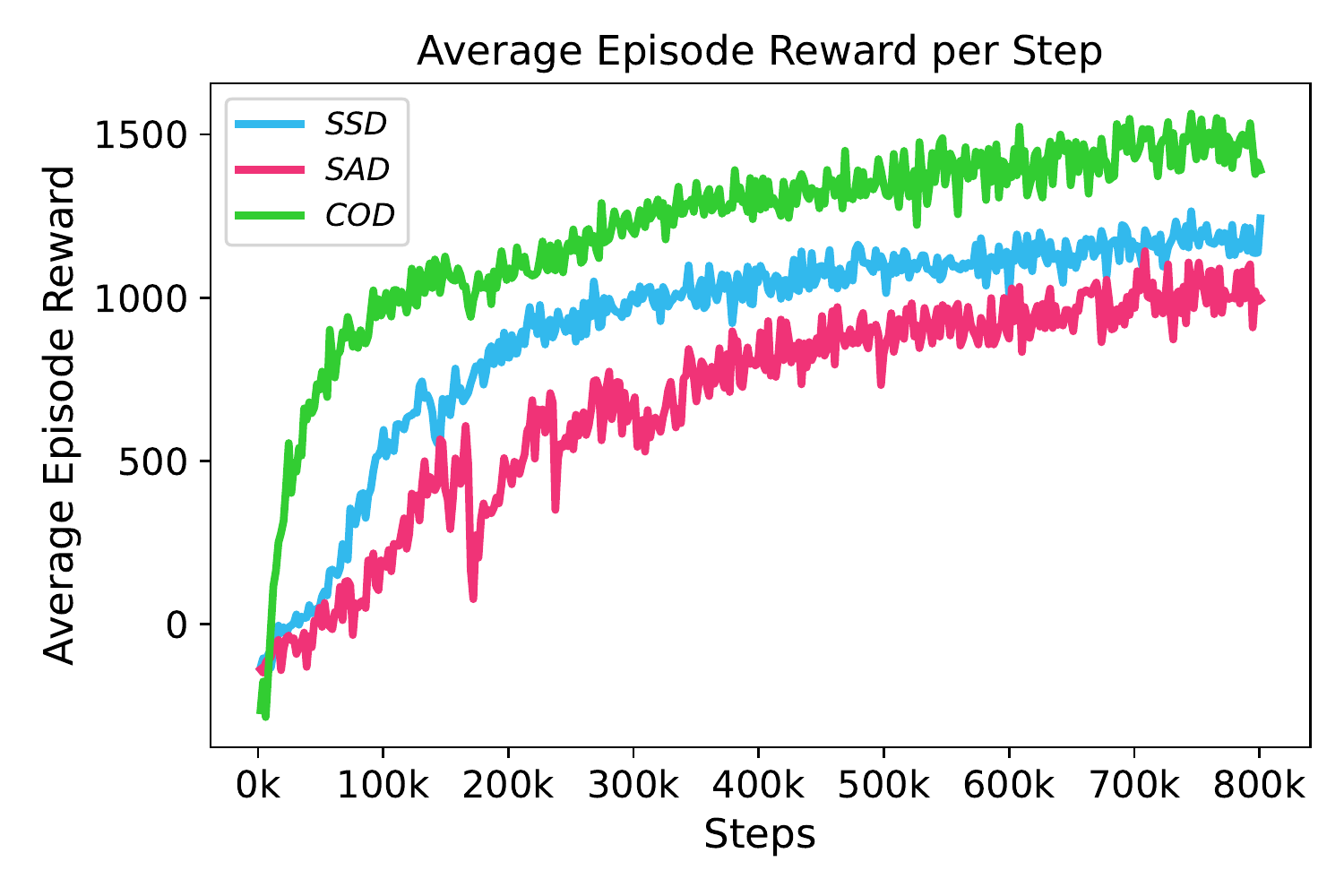}
         \caption{Average Episode Reward per Step in HT Detection for $c5315$}
         \label{fig:reward_C}
     \end{subfigure}
    \caption{The average episode length and reward vs. the number of steps in both HT insertion and detection for $c5315$}
    \vspace{-5mm}
    \label{fig:reward}
\end{figure*}

Figure~\ref{fig:reward} shows the average episode length and reward of the inserting and detector RL agents for the $c5315$ benchmark circuit. As seen from Figure~\ref{fig:reward}.a, initially,  the agent leans more towards ending the training episodes to avoid further losses. This trend continues until it gradually increases the episode length, increasing the reward, which can be observed in Figure~\ref{fig:reward}.b. Eventually, the agent collects more and more rewards. Although the agent accumulates higher rewards in $P_{high}$, the detection rate is not significantly different from $P_{rand}$. Figure~\ref{fig:reward}.c demonstrates the agent's ability to augment rewards in our three detection scenarios at an almost steady pace; it learns how to increase rewards along the way. It is worthwhile to point out that the proposed RL framework can save the state of the RL models at arbitrary intervals, which helps test the agent's efficacy at different timesteps. Note that since the detector's episode length is always $10$, this data was not included in the graph. The agent can always be trained for longer steps, but one should consider the trade-off between the time required and the accuracy achieved. 

\subsection{Test Vector Size vs. Accuracy}

\begin{figure}[!t]
\centering
     \includegraphics[width=0.49\textwidth]{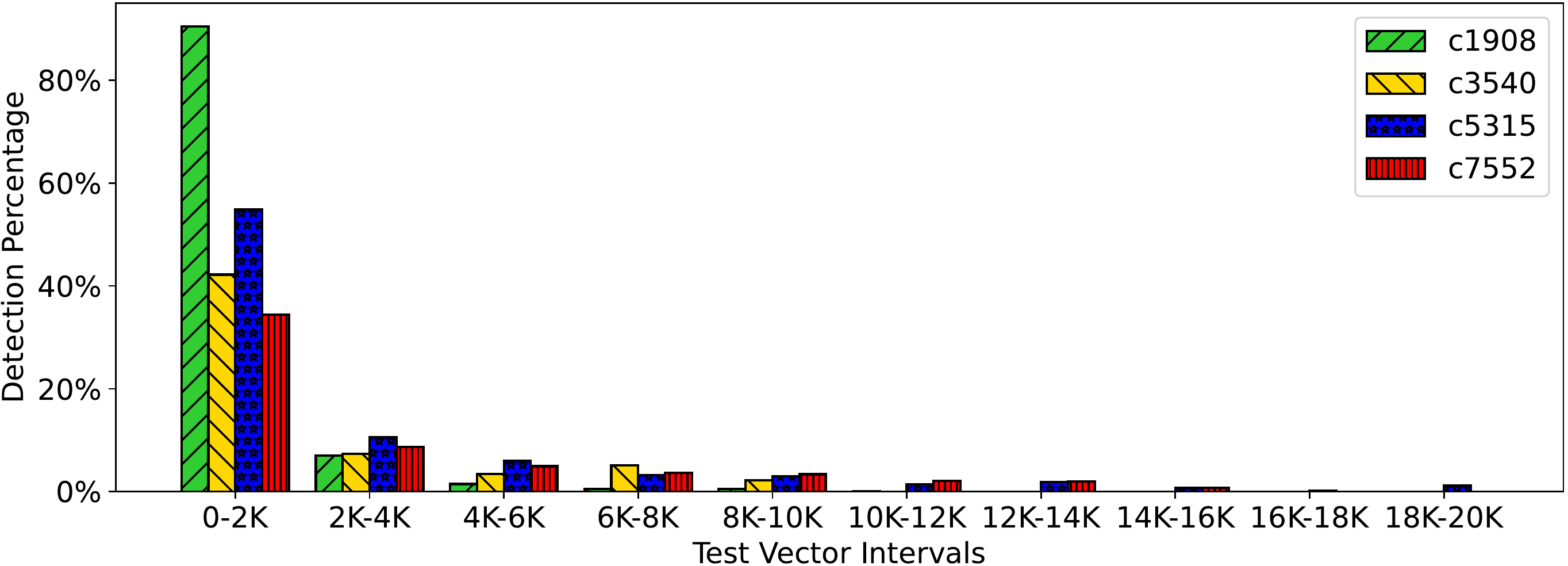}
    \caption{The number of generated test vectors (x-axis) vs. the HT detection accuracy (y-axis)}
    \vspace{-6mm}
    \label{fig:test_vector_length}
\end{figure}

We also investigate the relationship between the number of applied test vectors and the HT detection accuracy. For this experiment, we collect a set of test vectors that have obtained a certain minimum reward. We run the trained RL agent for $20$K episodes to identify such vectors. We set a cut-off reward of one-tenth of the collected reward in the last training episode. We collect $20$K test vectors that surpass this reward threshold. The HT detection distribution of the collected test vectors is shown in Figure~\ref{fig:test_vector_length} for $c1908$, $c3540$, $c5315$, and $c7552$ under the $P_{rand}$ insertion scenario and the $SAD$ detection scenario. 
The x-axis displays the intervals of the applied test vectors, and the y-axis shows the detection percentage of each particular interval. As can be seen, the first $2$K vectors have the greatest contribution toward HT detection. This figure is nearly $90\%$ for $c1908$ and just below $40\%$ for $c7552$. A similar comparison can be made between different HT detectors to help us find the relation between the quantity (number of test vectors) and the quality (the detection accuracy). Such analysis leads us to answer the question, ``Does adding more test vectors to the testing batch improve detection?" If the answer is negative, adopting more intelligent rewarding functions might be considered to offset this diminishing returns effect. That being said, in certain instances, adding more test batches leads to higher detection rates. We tested this scenario for $c3540$ where the \textit{Combined} detection rate with $20$K test patterns is around $80\%$ in the $P_{rand}$ scenario. We ran the trained detector agents $SSD$, $SAD$, and $COD$ for $20$K episodes, but this time, we collected all the test patterns that returned positive rewards. Accordingly, we collected $191$K, $183$K ,$121$K for $SSD$, $SAD$, and $COD$ and the detection rates were $89\%$, $86\%$, and $97\%$, respectively.


\subsection{RL Feasibility in Practice}
\changess{RL agents have been extensively used in various application domains where decision making is required, \eg, robotics  control~\cite{tai2017virtual,hwangbo2019learning}, gaming~\cite{silver2016mastering,silver2018general}, autonomous driving~\cite{talpaert2019exploring}, computer  architecture~\cite{zhou2019gdp,yin2018toward}, and hardware security~\cite{10031569}. Training RL models requires a large amount of interaction with the environment to learn an optimal policy. This can be costly in many environments (including the HT space) where the interactions with the environment are computationally expensive. The OpenAI RL agent that defeated the DOTA world champions famously took 10 months to train~\cite{berner2019dota}. Despite the training hurdle, RL introduces some valuable advantages in relationship to HTs.  First, RL facilitates the exploration of complex environments that humans cannot easily accomplish. It automates the decision-making process and eases automation especially where tasks must be performed repeatedly or in large volumes. RL, as an unsupervised learning technique, can build training sets for other agents that are then trained via supervised learning, for instance, an HT benchmark for training an ML-based HT detector. Moreover, RL removes the human bias stemming from a particular mindset in the process. RL has already proved to be a valuable solution in the HT domain~\cite{pan2021automated,gohil2022deterrent,gohil2022attrition}. While utilizing RL helps security engineers produce test vectors, the next generation of malicious actors might be bots designed to compromise security. Hence, despite the added layer of complexity, we believe that utilizing an RL approach for HT insertion and detection is feasible where the sheer complexity of the problem means that we need to explore all potential research avenues.}

\section{Conclusions and Future Directions}
\label{sec:Conc}
This paper presented the first framework for joint HT insertion and detection. The inserting and detection RL agents have tunable rewarding functions that enable researchers to experiment with different approaches to the problem. This framework will accelerate HT research by helping the research community evaluate their insertion/detection ideas with less effort. Our inserting tool provides a robust dataset that can be used for developing finer HT detectors, and our detector tool emphasizes the need for a multi-criteria detector that can cater to different HT insertion mindsets. We also presented a methodology to help the community compare HT detection methods, regardless of their implementation details. We applied this methodology to our HT detection and discovered that our tool offers the highest confidence in HT detection when using a combined detection scenario. We aim to explore more benchmarks and create a more diverse HT dataset for the community.

\changess{As an extension to this work, we aim to explore more benchmarks and provide support for other circuits, including sequential ones in our flow, \eg, ISCAS-89~\cite{vtWWWISCAS89} and ITC'99~\cite{utexasITCapos99Benchmark} benchmarks. One solution to tackle these circuits would be utilizing Design for Testability (DFT) techniques such as scan chains~\cite{narayanan1993optimal}. In a full scan design, memory elements are connected to the chain, enabling test engineers to use combinational test patterns instead of sequential ones. Given the existence of this playground infrastructure, further research questions and more complex ideas can be explored.
}

\section*{Declarations}

\subsection*{Funding}
This work has been partially funded by NSF grants 2219680 and 2219679. 
\vspace{-3mm}
\subsection*{Availability of Data and Materials}
Our HT benchmark and test vectors are available on \url{https://github.com/NMSU-PEARL/Hardware-Trojan-Insertion-and-Detection-with-Reinforcement-Learning}
We will share the hardware Trojan benchmark on a case-by-case basis. Requests can be made on the same GitHub repository.


\bibliographystyle{References-Style/IEEEtran}
\bibliography{References-Style/IEEEabrv, References-Style/IEEEexample}

\begin{thebibliography}{10}
\providecommand{\url}[1]{#1}
\csname url@samestyle\endcsname
\providecommand{\newblock}{\relax}
\providecommand{\bibinfo}[2]{#2}
\providecommand{\BIBentrySTDinterwordspacing}{\spaceskip=0pt\relax}
\providecommand{\BIBentryALTinterwordstretchfactor}{4}
\providecommand{\BIBentryALTinterwordspacing}{\spaceskip=\fontdimen2\font plus
\BIBentryALTinterwordstretchfactor\fontdimen3\font minus
  \fontdimen4\font\relax}
\providecommand{\BIBforeignlanguage}[2]{{%
\expandafter\ifx\csname l@#1\endcsname\relax
\typeout{** WARNING: IEEEtran.bst: No hyphenation pattern has been}%
\typeout{** loaded for the language `#1'. Using the pattern for}%
\typeout{** the default language instead.}%
\else
\language=\csname l@#1\endcsname
\fi
#2}}
\providecommand{\BIBdecl}{\relax}
\BIBdecl

\bibitem{securing}
\BIBentryALTinterwordspacing
``Securing defense-critical supply chains: An action plan developed in response
  to president biden's executive order 14017.'' [Online]. Available:
  \url{https://tinyurl.com/3wmddx5d}
\BIBentrySTDinterwordspacing

\bibitem{pan2021automated}
Z.~Pan and P.~Mishra, ``Automated test generation for hardware trojan detection
  using reinforcement learning,'' in \emph{Proceedings of the 26th Asia and
  South Pacific Design Automation Conference}, 2021, pp. 408--413.

\bibitem{shakya2017benchmarking}
B.~Shakya, T.~He, H.~Salmani, D.~Forte, S.~Bhunia, and M.~Tehranipoor,
  ``Benchmarking of hardware trojans and maliciously affected circuits,''
  \emph{Journal of Hardware and Systems Security}, vol.~1, no.~1, pp. 85--102,
  2017.

\bibitem{sarihi2021survey}
A.~Sarihi, A.~Patooghy, A.~Khalid, M.~Hasanzadeh, M.~Said, and A.-H.~A. Badawy,
  ``A survey on the security of wired, wireless, and 3d network-on-chips,''
  \emph{IEEE Access}, 2021.

\bibitem{salmani2013design}
H.~Salmani, M.~Tehranipoor, and R.~Karri, ``On design vulnerability analysis
  and trust benchmarks development,'' in \emph{2013 IEEE 31st international
  conference on computer design (ICCD)}.\hskip 1em plus 0.5em minus 0.4em\relax
  IEEE, 2013, pp. 471--474.

\bibitem{trusthub}
``{Trust-Hub},'' \url{https://trust-hub.org/}, accessed: 2023-11-08.

\bibitem{salmani2016cotd}
H.~Salmani, ``Cotd: Reference-free hardware trojan detection and recovery based
  on controllability and observability in gate-level netlist,'' \emph{IEEE
  Transactions on Information Forensics and Security}, vol.~12, no.~2, pp.
  338--350, 2016.

\bibitem{hasegawa2017trojan}
K.~Hasegawa, M.~Yanagisawa, and N.~Togawa, ``Trojan-feature extraction at
  gate-level netlists and its application to hardware-trojan detection using
  random forest classifier,'' in \emph{2017 IEEE International Symposium on
  Circuits and Systems (ISCAS)}.\hskip 1em plus 0.5em minus 0.4em\relax IEEE,
  2017, pp. 1--4.

\bibitem{sebt2018circuit}
S.~M. Sebt, A.~Patooghy, H.~Beitollahi, and M.~Kinsy, ``Circuit enclaves
  susceptible to hardware trojans insertion at gate-level designs,'' \emph{IET
  Computers \& Digital Techniques}, vol.~12, no.~6, pp. 251--257, 2018.

\bibitem{gohil2022deterrent}
V.~Gohil, S.~Patnaik, H.~Guo, D.~Kalathil, and J.~Rajendran, ``Deterrent:
  detecting trojans using reinforcement learning,'' in \emph{Proceedings of the
  59th ACM/IEEE Design Automation Conference}, 2022, pp. 697--702.

\bibitem{cruz2018automated}
J.~Cruz, Y.~Huang, P.~Mishra, and S.~Bhunia, ``An automated configurable trojan
  insertion framework for dynamic trust benchmarks,'' in \emph{2018 Design,
  Automation \& Test in Europe Conference \& Exhibition (DATE)}.\hskip 1em plus
  0.5em minus 0.4em\relax IEEE, 2018, pp. 1598--1603.

\bibitem{lyu2020scalable}
Y.~Lyu and P.~Mishra, ``Scalable activation of rare triggers in hardware
  trojans by repeated maximal clique sampling,'' \emph{IEEE Transactions on
  Computer-Aided Design of Integrated Circuits and Systems}, vol.~40, no.~7,
  pp. 1287--1300, 2020.

\bibitem{2018FyrbiakHAL}
M.~Fyrbiak, S.~Wallat, P.~Swierczynski, M.~Hoffmann, S.~Hoppach, M.~Wilhelm,
  T.~Weidlich, R.~Tessier, and C.~Paar, ``{HAL-} the missing piece of the
  puzzle for hardware reverse engineering, trojan detection and insertion,''
  \emph{IEEE Transactions on Dependable and Secure Computing}, 2018.

\bibitem{yu2019improved}
S.~Yu, W.~Liu, and M.~O'Neill, ``An improved automatic hardware trojan
  generation platform,'' in \emph{2019 IEEE Computer Society Annual Symposium
  on VLSI (ISVLSI)}.\hskip 1em plus 0.5em minus 0.4em\relax IEEE, 2019, pp.
  302--307.

\bibitem{chakraborty2009mero}
R.~S. Chakraborty, F.~Wolff, S.~Paul, C.~Papachristou, and S.~Bhunia, ``Mero: A
  statistical approach for hardware trojan detection,'' in \emph{International
  Workshop on Cryptographic Hardware and Embedded Systems}.\hskip 1em plus
  0.5em minus 0.4em\relax Springer, 2009, pp. 396--410.

\bibitem{gohil2022attrition}
V.~Gohil, H.~Guo, S.~Patnaik, and J.~Rajendran, ``Attrition: Attacking static
  hardware trojan detection techniques using reinforcement learning,'' in
  \emph{Proceedings of the 2022 ACM SIGSAC Conference on Computer and
  Communications Security}, 2022, pp. 1275--1289.

\bibitem{schulman2017proximal}
J.~Schulman, F.~Wolski, P.~Dhariwal, A.~Radford, and O.~Klimov, ``Proximal
  policy optimization algorithms,'' \emph{arXiv preprint arXiv:1707.06347},
  2017.

\bibitem{sarihi2022hardware}
A.~Sarihi, A.~Patooghy, P.~Jamieson, and A.-H.~A. Badawy, ``Hardware trojan
  insertion using reinforcement learning,'' in \emph{Proceedings of the Great
  Lakes Symposium on VLSI 2022}, 2022, pp. 139--142.

\bibitem{sarihi2023multi}
A.~Sarihi, P.~Jamieson, A.~Patooghy, and A.-H.~A. Badawy, ``Multi-criteria
  hardware trojan detection: A reinforcement learning approach,'' \emph{arXiv
  preprint arXiv:2304.13232}, 2023.

\bibitem{krieg2023reflections}
C.~Krieg, ``Reflections on trusting trusthub,'' in \emph{2023 IEEE/ACM
  International Conference on Computer Aided Design (ICCAD)}.\hskip 1em plus
  0.5em minus 0.4em\relax IEEE, 2023, pp. 1--9.

\bibitem{jyothi2017taint}
V.~Jyothi, P.~Krishnamurthy, F.~Khorrami, and R.~Karri, ``Taint: Tool for
  automated insertion of trojans,'' in \emph{2017 IEEE International Conference
  on Computer Design (ICCD)}.\hskip 1em plus 0.5em minus 0.4em\relax IEEE,
  2017, pp. 545--548.

\bibitem{wallat2017look}
S.~Wallat, M.~Fyrbiak, M.~Schl{\"o}gel, and C.~Paar, ``A look at the dark side
  of hardware reverse engineering-a case study,'' in \emph{2017 IEEE 2nd
  International Verification and Security Workshop (IVSW)}.\hskip 1em plus
  0.5em minus 0.4em\relax IEEE, 2017, pp. 95--100.

\bibitem{cruz2022automatic}
J.~Cruz, P.~Gaikwad, A.~Nair, P.~Chakraborty, and S.~Bhunia, ``Automatic
  hardware trojan insertion using machine learning,'' \emph{arXiv preprint
  arXiv:2204.08580}, 2022.

\bibitem{perez2021side}
T.~Perez, M.~Imran, P.~Vaz, and S.~Pagliarini, ``Side-channel trojan
  insertion-a practical foundry-side attack via eco,'' in \emph{2021 IEEE
  International Symposium on Circuits and Systems (ISCAS)}.\hskip 1em plus
  0.5em minus 0.4em\relax IEEE, 2021, pp. 1--5.

\bibitem{perez2022hardware}
T.~Perez and S.~Pagliarini, ``Hardware trojan insertion in finalized layouts:
  From methodology to a silicon demonstration,'' \emph{IEEE Transactions on
  Computer-Aided Design of Integrated Circuits and Systems}, 2022.

\bibitem{puschner2023red}
E.~Puschner, T.~Moos, S.~Becker, C.~Kison, A.~Moradi, and C.~Paar, ``Red team
  vs. blue team: A real-world hardware trojan detection case study across four
  modern cmos technology generations,'' in \emph{2023 IEEE Symposium on
  Security and Privacy (SP)}.\hskip 1em plus 0.5em minus 0.4em\relax IEEE,
  2023, pp. 56--74.

\bibitem{hepp2022pragmatic}
A.~Hepp, T.~Perez, S.~Pagliarini, and G.~Sigl, ``A pragmatic methodology for
  blind hardware trojan insertion in finalized layouts,'' in \emph{Proceedings
  of the 41st IEEE/ACM International Conference on Computer-Aided Design},
  2022, pp. 1--9.

\bibitem{nozawa2021generating}
K.~Nozawa, K.~Hasegawa, S.~Hidano, S.~Kiyomoto, K.~Hashimoto, and N.~Togawa,
  ``Generating adversarial examples for hardware-trojan detection at gate-level
  netlists,'' \emph{Journal of Information Processing}, vol.~29, pp. 236--246,
  2021.

\bibitem{goldstein1980scoap}
L.~H. Goldstein and E.~L. Thigpen, ``Scoap: Sandia
  controllability/observability analysis program,'' in \emph{Proceedings of the
  17th Design Automation Conference}, 1980, pp. 190--196.

\bibitem{yu2021hw2vec}
S.-Y. Yu, R.~Yasaei, Q.~Zhou, T.~Nguyen, and M.~A. Al~Faruque, ``Hw2vec: A
  graph learning tool for automating hardware security,'' in \emph{2021 IEEE
  International Symposium on Hardware Oriented Security and Trust
  (HOST)}.\hskip 1em plus 0.5em minus 0.4em\relax IEEE, 2021, pp. 13--23.

\bibitem{wolf2013yosys}
C.~Wolf, J.~Glaser, and J.~Kepler, ``Yosys-a free verilog synthesis suite,'' in
  \emph{Proceedings of the 21st Austrian Workshop on Microelectronics
  (Austrochip)}, 2013.

\bibitem{bassett2015introduction}
\BIBentryALTinterwordspacing
L.~Bassett, \emph{Introduction to JavaScript Object Notation: A To-the-Point
  Guide to JSON}.\hskip 1em plus 0.5em minus 0.4em\relax Sebastopol: O'Reilly
  Media, 2015. [Online]. Available:
  \url{https://books.google.com/books?id=Qv9PCgAAQBAJ}
\BIBentrySTDinterwordspacing

\bibitem{DBLP:journals/corr/BrockmanCPSSTZ16}
\BIBentryALTinterwordspacing
G.~Brockman, V.~Cheung, L.~Pettersson, J.~Schneider, J.~Schulman, J.~Tang, and
  W.~Zaremba, ``Openai gym,'' \emph{CoRR}, vol. abs/1606.01540, 2016. [Online].
  Available: \url{http://arxiv.org/abs/1606.01540}
\BIBentrySTDinterwordspacing

\bibitem{bushnell2000essentials}
M.~L. Bushnell, ``Essentials of electronic testing for digital,'' \emph{Memory
  \& Mixed-Signal VLSI Circuits}, 2000.

\bibitem{nguyen2019deep}
T.~T. Nguyen and V.~J. Reddi, ``Deep reinforcement learning for cyber
  security,'' \emph{IEEE Transactions on Neural Networks and Learning Systems},
  2019.

\bibitem{raffin2019stable}
\BIBentryALTinterwordspacing
A.~Raffin, A.~Hill, A.~Gleave, A.~Kanervisto, M.~Ernestus, and N.~Dormann,
  ``Stable-baselines3: Reliable reinforcement learning implementations,''
  \emph{Journal of Machine Learning Research}, vol.~22, no. 268, pp. 1--8,
  2021. [Online]. Available: \url{http://jmlr.org/papers/v22/20-1364.html}
\BIBentrySTDinterwordspacing

\bibitem{bryan1985iscas}
D.~Bryan, ``{The {ISCAS'85} benchmark circuits and netlist format},'' p.~39,
  1985.

\bibitem{SciPyProceedings_11}
A.~A. Hagberg, D.~A. Schult, and P.~J. Swart, ``Exploring network structure,
  dynamics, and function using networkx,'' in \emph{Proceedings of the 7th
  Python in Science Conference}, G.~Varoquaux, T.~Vaught, and J.~Millman, Eds.,
  Pasadena, CA USA, 2008, pp. 11 -- 15.

\bibitem{githubGitHubNMSUPEARLHardwareTrojanInsertionandDetectionwithReinforcementLearning}
``{G}it{H}ub -
  {N}{M}{S}{U}-{P}{E}{A}{R}{L}/{H}ardware-{T}rojan-{I}nsertion-and-{D}etection-with-{R}einforcement-{L}earning:
  {R}einforcement {L}earning-based {H}ardware {T}rojan {D}etector ---
  github.com,''
  \url{https://github.com/NMSU-PEARL/Hardware-Trojan-Insertion-and-Detection-with-Reinforcement-Learning},
  [Accessed 27-12-2023].

\bibitem{iscas85}
``{ISCAS High-Level Models},''
  \url{https://web.eecs.umich.edu/~jhayes/iscas.restore/benchmark.html},
  accessed: 2023-11-07.

\bibitem{amaru2015epfl}
L.~Amar{\'u}, P.-E. Gaillardon, and G.~De~Micheli, ``The epfl combinational
  benchmark suite,'' in \emph{Proceedings of the 24th International Workshop on
  Logic \& Synthesis (IWLS)}, no. CONF, 2015.

\bibitem{tai2017virtual}
L.~Tai, G.~Paolo, and M.~Liu, ``Virtual-to-real deep reinforcement learning:
  Continuous control of mobile robots for mapless navigation,'' in \emph{2017
  IEEE/RSJ International Conference on Intelligent Robots and Systems
  (IROS)}.\hskip 1em plus 0.5em minus 0.4em\relax IEEE, 2017, pp. 31--36.

\bibitem{hwangbo2019learning}
J.~Hwangbo, J.~Lee, A.~Dosovitskiy, D.~Bellicoso, V.~Tsounis, V.~Koltun, and
  M.~Hutter, ``Learning agile and dynamic motor skills for legged robots,''
  \emph{Science Robotics}, vol.~4, no.~26, p. eaau5872, 2019.

\bibitem{silver2016mastering}
D.~Silver, A.~Huang, C.~J. Maddison, A.~Guez, L.~Sifre, G.~Van Den~Driessche,
  J.~Schrittwieser, I.~Antonoglou, V.~Panneershelvam, M.~Lanctot \emph{et~al.},
  ``Mastering the game of go with deep neural networks and tree search,''
  \emph{nature}, vol. 529, no. 7587, pp. 484--489, 2016.

\bibitem{silver2018general}
D.~Silver, T.~Hubert, J.~Schrittwieser, I.~Antonoglou, M.~Lai, A.~Guez,
  M.~Lanctot, L.~Sifre, D.~Kumaran, T.~Graepel \emph{et~al.}, ``A general
  reinforcement learning algorithm that masters chess, shogi, and go through
  self-play,'' \emph{Science}, vol. 362, no. 6419, pp. 1140--1144, 2018.

\bibitem{talpaert2019exploring}
V.~Talpaert, I.~Sobh, B.~R. Kiran, P.~Mannion, S.~Yogamani, A.~El-Sallab, and
  P.~Perez, ``Exploring applications of deep reinforcement learning for
  real-world autonomous driving systems,'' \emph{arXiv preprint
  arXiv:1901.01536}, 2019.

\bibitem{zhou2019gdp}
Y.~Zhou, S.~Roy, A.~Abdolrashidi, D.~Wong, P.~C. Ma, Q.~Xu, M.~Zhong, H.~Liu,
  A.~Goldie, A.~Mirhoseini \emph{et~al.}, ``Gdp: Generalized device placement
  for dataflow graphs,'' \emph{arXiv preprint arXiv:1910.01578}, 2019.

\bibitem{yin2018toward}
J.~Yin, Y.~Eckert, S.~Che, M.~Oskin, and G.~H. Loh, ``Toward more efficient noc
  arbitration: A deep reinforcement learning approach,'' in \emph{Proc. IEEE
  1st Int. Workshop AI-assisted Des. Architecture}, vol. 128, 2018.

\bibitem{10031569}
S.~Patnaik, V.~Gohil, H.~Guo, and J.~J. Rajendran, ``Reinforcement learning for
  hardware security: Opportunities, developments, and challenges,'' in
  \emph{2022 19th International SoC Design Conference (ISOCC)}, 2022, pp.
  217--218.

\bibitem{berner2019dota}
C.~Berner, G.~Brockman, B.~Chan, V.~Cheung, P.~Debiak, C.~Dennison, D.~Farhi,
  Q.~Fischer, S.~Hashme, C.~Hesse \emph{et~al.}, ``Dota 2 with large scale deep
  reinforcement learning,'' \emph{arXiv preprint arXiv:1912.06680}, 2019.

\bibitem{vtWWWISCAS89}
``{W}{W}{W}: {I}{S}{C}{A}{S}89 {S}equential {B}enchmark {C}ircuits ---
  filebox.ece.vt.edu,'' \url{https://filebox.ece.vt.edu/~mhsiao/iscas89.html},
  [Accessed 22-01-2024].

\bibitem{utexasITCapos99Benchmark}
``{I}{T}{C}'99 {B}enchmark {H}omepage --- cerc.utexas.edu,''
  \url{https://www.cerc.utexas.edu/itc99-benchmarks/bench.html}, [Accessed
  22-01-2024].

\bibitem{narayanan1993optimal}
S.~Narayanan, R.~Gupta, and M.~A. Breuer, ``Optimal configuring of multiple
  scan chains,'' \emph{IEEE transactions on computers}, vol.~42, no.~9, pp.
  1121--1131, 1993.

\end{thebibliography}
\vspace{-13mm}
\begin{IEEEbiography}[{\includegraphics[width=1in,height=1.25in, clip,keepaspectratio]{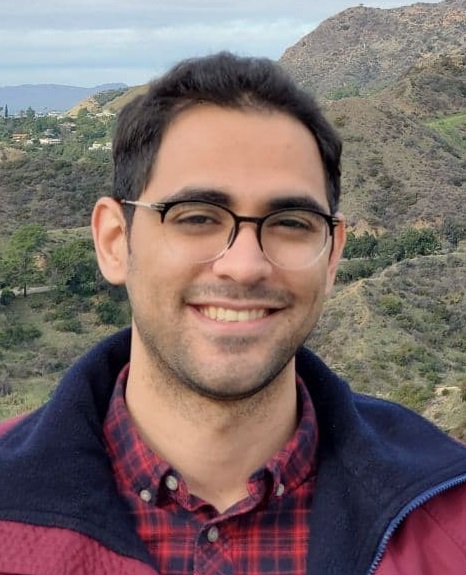}}]{Amin Sarihi} received his B.Sc.\ degree in Electrical Engineering at Shiraz University, Shiraz, Iran, in 2015, and an M.Sc.\ degree in Electrical Engineering, in 2019, focusing on digital systems at Iran University Science and Technology, Tehran, Iran. He is a Ph.D.\ candidate in Electrical and Computer Engineering at New Mexico State University, NM, USA. His research interest includes hardware security and trust, network-on-chip security, embedded systems security, machine learning, and computer performance evaluation.  
\end{IEEEbiography}
\vspace{-12mm}

\begin{IEEEbiography}[{\includegraphics[width=1in,height=1.25in, clip,keepaspectratio]{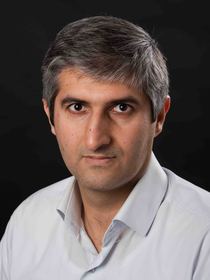}}]{Ahmad Patooghy} received his Ph.D.\ degree in Computer Engineering from Sharif University of Technology, Tehran, Iran, in 2011. From 2011 to 2017, he was an Assistant Professor at Iran University of Science and Technology in Tehran. Then, he joined the University of Central Arkansas, AR, USA (2018 to 2021). Since August 2021, he has been with the Department of Computer Systems Technology at North Carolina Agricultural and Technical State University, leading the Intelligent and Embedded Systems Laboratory. He has published more than 100 journal papers and conference proceedings. He has been a reviewer and PC member for various IEEE, ACM, Elsevier, and Springer journals and conferences. He has also served as a panelist for the National Science Foundation for reviewing grant proposals. His research interests include the security and reliability of machine learning applications and accelerators, hardware security in IoT and cyber-physical systems, and hardware design for machine learning acceleration. Dr. Patooghy is a Senior Member of the IEEE, IEEE Computer Society, and a Member of the ACM.  
\end{IEEEbiography}

\begin{IEEEbiography}[{\includegraphics[width=1in,height=1.25in, clip,keepaspectratio]{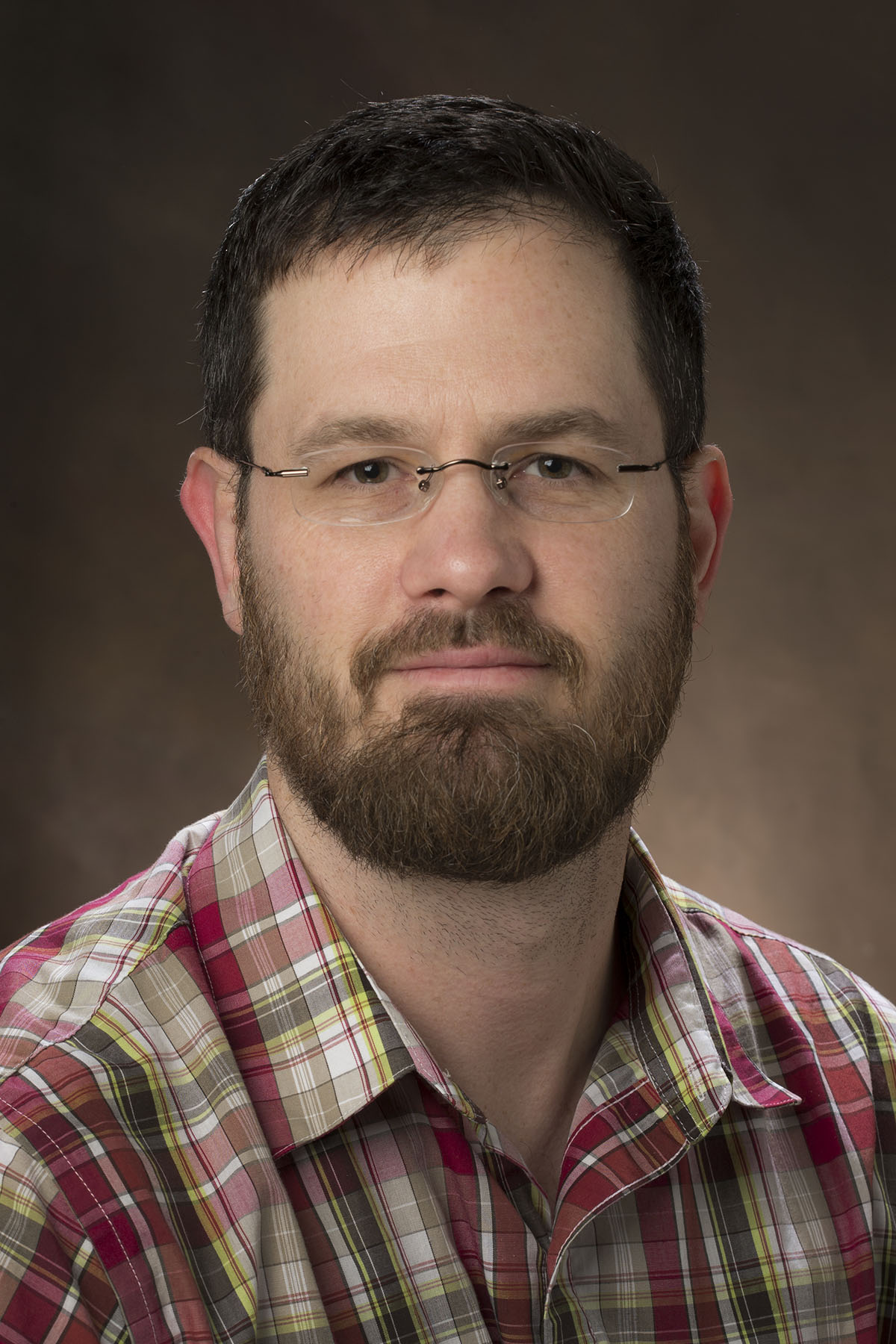}}]{Peter Jamieson}
is an Associate Professor at Miami University in the Electrical and Computer Engineering department. Dr.\  Peter Jamieson's research focuses broadly on computation, education, games, and the intersections between these domains. He received his Ph.D.\ from the University of Toronto in 2007 and had post-doc positions at both Toronto and Imperial College before joining the faculty in Electrical and Computer Engineering at Miami University in 2009.
\end{IEEEbiography}
\vspace{-128mm}

\begin{IEEEbiography}[{\includegraphics[width=1.0in,height=1.25in, clip,keepaspectratio]{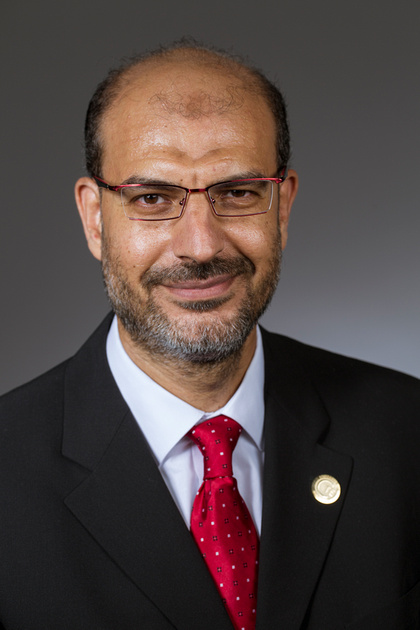}}]{Abdel-Hameed Badawy} received his Ph.D.\ and M.Sc.\ from the University of Maryland, both in 2013 and 2002 respectively, in Computer Engineering. He obtained his B.Sc. in Electronics Engineering with a concentration on Computers \& Control Systems from Mansoura University, Egypt, where he was ranked first in his graduating class in 1996. He is an associate professor in the Klipsch School of Electrical and Computer Engineering at New Mexico State University, Las Cruces, NM. He is also a Los Alamos Joint Faculty with the New Mexico Consortium. He has been a visiting research scientist at the New Mexico Consortium. He was a lead research scientist at the High-Performance Computing Laboratory at George Washington University. His research interests include computer architecture, high-performance computing, performance modeling and prediction, green computing, hardware security, and optical computing. He has more than 80 publications in conferences and Journals. He has been a PC member and reviewer for various IEEE, ACM, Elsevier, and Springer journals. He has also served as a panelist at the National Science Foundation and Department of Energy for reviewing grant proposals. His research has been awarded the best paper/poster awards. He is a Senior Member of the IEEE, IEEE Computer Society, a Professional Member of the ACM, and a Dean\textquotesingle s member of the ASEE. He served as the vice-chair of the Arkansas River Valley IEEE section in 2014 and 2016.
\end{IEEEbiography}




\end{document}